\documentclass[sigplan,screen]{acmart}
\usepackage{graphicx}
\usepackage{subfigure}
\usepackage{multirow}
\usepackage{float}
\usepackage{algorithm,algpseudocode,amsmath}
\renewcommand\footnotetextcopyrightpermission[1]{} 
\newcommand{\scheduler}{Dynamic Asymmetry Scheduler}

\makeatletter 

\let\wfs@comment@comment\comment
\let\comment\@undefined

\usepackage[final]{changes}

\definechangesauthor[color=teal]{MP}
\definechangesauthor[color=darkgray]{MM}
\definechangesauthor[color=blue]{MA}
\definechangesauthor[color=red]{JC}
\definechangesauthor[color=cyan]{PN}

\makeatother

\title{Scheduling Task-parallel Applications in Dynamically Asymmetric Environments}

 \author{Jing Chen}
 \affiliation{\institution{Chalmers University of Technology}}
 \email{chjing@chalmers.se}
 \author{Pirah Noor Soomro}
 \affiliation{\institution{Chalmers University of Technology}}
 \email{pirah@chalmers.se}
  \author{Mustafa Abduljabbar}
 \affiliation{\institution{Chalmers University of Technology}}
 \email{musabdu@chalmers.se}
  \author{Madhavan Manivannan}
 \affiliation{\institution{Chalmers University of Technology}}
 \email{madhavan@chalmers.se}
  \author{Miquel Peric\`as}
 \affiliation{\institution{Chalmers University of Technology}}
 \email{miquelp@chalmers.se}

\newif\ifdonotcomment
\donotcommentfalse

\newcommand{\testcomment}[1]{
\ifdonotcomment
\textcolor{red}{#1}
\fi
}



\begin{abstract}
Shared resource interference is observed by applications as dynamic performance asymmetry.
Prior art has developed approaches to reduce the impact of performance asymmetry mainly at the operating system and architectural levels. 
In this work, we study how application-level scheduling techniques can leverage \replaced[id=MP]{moldability }{knowledge of task parallelism}\added[id=MA]{(i.e. flexibility to work as either single-threaded or multithreaded task)} and \added[id=MP]{explicit knowledge on} task criticality to 
handle scenarios in which system performance is not only unknown but also changing over time.
Our proposed task scheduler dynamically learns the performance characteristics of the underlying platform and uses this knowledge to 
devise better schedules aware of dynamic performance asymmetry, hence reducing the impact of interference. 
Our evaluation shows that both criticality-aware scheduling and parallelism tuning are effective schemes to address interference in both shared and distributed memory applications. 

\end{abstract}

\begin{document}
\setlength{\marginparwidth}{2cm}
\settopmatter{printfolios=true}
\maketitle
\pagestyle{plain}
\renewcommand{\shortauthors}{J. Chen, P. N. Soomro, M. Abduljabbar, M. Manivannan, and M. Pericàs}

\section{Introduction}
Modern computer systems are designed to handle a large variety of dynamic events triggered by multiple sources, such as I/O, user activity, scheduled jobs, O/S, power management, etc. 
Applications sharing system resources will observe a performance \added[id=MP]{(i.e.~execution time)} \replaced[id=MM]{variability}{variation} during such \added[id=MM]{interfering} events. \deleted[id=MM]{, also known as \textit{interference}} 
Understanding the causes and properties of interference is thus of great importance. 
Prior work has mainly focused on HPC environments, where execution time delays are particularly costly~\cite{hoefler-sc10,ates-icpp19,skinner-wcs05}. 
In these scenarios, common strategies to reduce interference include eliminating unnecessary activities such as process scheduling or memory management~\cite{moreira-sc06}, and carefully managing interrupts to minimize application perturbation~\cite{riesen-cc09}. 
As the scale and heterogeneity of systems keeps increasing, tackling interference becomes an even larger concern~\cite{patki-sc19}.
While the aforementioned techniques can generally mitigate interference, sources of performance variability are overall very diverse in modern highly interconnected systems, and can only be addressed effectively in cooperation with application knowledge.

Applications are typically not aware of interference, but rather observe such events as temporary episodes of performance asymmetry. Performance asymmetry can be defined as the case when individual cores have different \replaced[id=MP]{progress rates (e.g.~MIPS)}{performance}~\cite{balakrishnan-isca06}. 
From an architectural perspective, sources of asymmetry can be broadly categorized into fixed and dynamic. 
Fixed sources appear due to hardware features, for example, same-ISA cores with different compute capabilities (e.g.~big.LITTLE~\cite{arm-big-little}). 
Hence, fixed asymmetry is permanent and does not evolve over time.
Dynamic sources of performance asymmetry arise from execution-time activities such as DVFS for power management~\cite{lesueur-hotpower10} or the sharing of resources between applications~\cite{zhuravlev-asplos10}. 

Performance asymmetry is particularly problematic for multithreaded applications with frequent synchronization.
A simple event slowing down the execution of a single thread can potentially result in a global performance degradation by delaying sibling threads waiting at a synchronization point. 
Unlike related work that has focused on system-level approaches, this paper explores runtime-level techniques applicable to DAG-based parallel applications. 
\added[id=MP]{Runtime systems usually operate at the user level and have little control over the system. Hence, one of the scenarios that we consider is DVFS activity that is beyond control of the runtime system}. \added[id=MM]{The second scenario that we consider are applications that are co-scheduled for execution, as is generally the case in HPC environments and in data centers.}
Scheduling such applications in scenarios with fixed performance asymmetry has been studied recently. For example, CATS~\cite{chronaki-ics15} and GCA~\cite{hsu-ica3pp07} are schedulers that \replaced[id=MA]{use algorithms to identify the critical path }{exploit a notion of task's criticality} to prioritize tasks and schedule them on the faster cores. However, \deleted[id=MM]{even} if faster cores suffer from interference, these schedulers will continue placing critical tasks on the perturbed partitions \added[id=MA]{(a set of execution resources such as cores or sockets)}, potentially leading to sub-optimal performance.
Continuous introspection is a step towards handling more dynamic environments. The AllScale runtime~\cite{allscale} collects runtime performance information and feeds it into an optimiser module that can tune thread numbers and use DVFS to optimize the execution. However, this runtime targets homogeneous architectures, and the solution is not applicable to systems exhibiting dynamic per-core performance asymmetry (such as interference).  
The high performance scheduling of DAGs on multicore systems in which cores are not only performance asymmetric but have performance that varies at runtime is -to the best of our knowledge- an unsolved problem in literature.

Our study indicates that statically mapping critical tasks to a fixed core results in suboptimal performance in the presence of \deleted[id=MM]{background  activities that cause} dynamic performance asymmetry. 
Our first observation is that a performance model that dynamically learns and updates the platform's performance characteristics needs to be developed to quickly yet consistently identify interference. In this paper, we leverage a simple tracing scheme called Performance Trace Table (PTT)~\cite{rohlin-hip3es19} that can be used to predict the performance characteristics on a per-task level. 
Next, we develop a task scheduler targeting systems with unknown and variable performance characteristics. 
Starting from a fixed-asymmetry scheduling baseline similar to CATS, we explore two novel directions. In the first approach, we leverage the PTT to steer critical tasks to the higher performing cores, in an attempt to speedup the critical tasks. In the second approach, we use the performance model (PTT) to choose an appropriate partition of cores for each task. This is done to reduce interference resulting from resource oversubscription.  
The aforementioned approaches are then evaluated in two \replaced[id=MM]{specific}{common implementation of the} interference scenarios: dynamic power management and job co-scheduling. \added[id=MM]{These scenarios occur frequently in HPC environments and in data centers due to techniques employed for improving utilization and energy efficiency under power limits.}
While aggregating cores to reduce oversubscription provides improved robustness to interference, our findings indicate that steering critical tasks provides higher gains in the presence of interfering activities and should be thus prioritized, unless the workload cannot differentiate between critical and non-critical tasks.  

In summary, this paper proposes application-level scheduling techniques to improve interference-robustness in parallel applications. The main contributions of this work are as follows:

\begin{itemize}
\item We explore the performance of \deleted[id=MP]{traditional} random work stealing and modern fixed-asymmetry criticality-based schedulers in the presence of interference and observe that the achieved performance is largely suboptimal. 
\item We show how simple online trace-based models can be used to model the platform's performance characteristics on a per-task basis, including the case in which the per-core performance characteristics vary over time.  
\item We explore two techniques to reduce the impact of interference.
First, we analyze the impact of steering critical tasks to cores dynamically detected as being fastest according to the trace-based model.
Second, we study the impact of assigning multi-core partitions for the execution of each individual task. 
\end{itemize}

The remainder of this paper is organized as follows. 
In Section~\ref{background} we describe the execution model that underlies this research. 
We present our runtime approach in Section~\ref{approach}. 
Section~\ref{setup} describes the experimental setup, including the implementation and the benchmarks which are used for the evaluation of our approach in Section~\ref{Perf_Eva}.
Finally, Section~\ref{RelatedWork} describes related work, while Section~\ref{conclusion} concludes the work. 
\section{Background} 
\label{background}

We consider DAGs composed of tasks having \replaced[id=MP]{either high or low}{different} \textit{priority}. 
DAGs are commonly used to describe the computations of multithreaded applications~\cite{openmp45-api,duran-ppl11,blumofe-jacm99} and of workflows composed of multiple dependent programs~\cite{ludascher-cc06}.  
The DAG model is a very general model for program execution that can support both regular and irregular computations. Regular computations can be described by creating all nodes and edges ahead of program execution (static DAG). Iterative programs with intra-loop parallelism can be described by unrolling the external loop and generating a layer of tasks for each iteration. In addition, irregular computations can be described by allowing tasks to conditionally insert new tasks into the DAG at runtime (dynamic DAG). 

\begin{figure}[t]
\centering
\includegraphics[width=0.55\columnwidth]{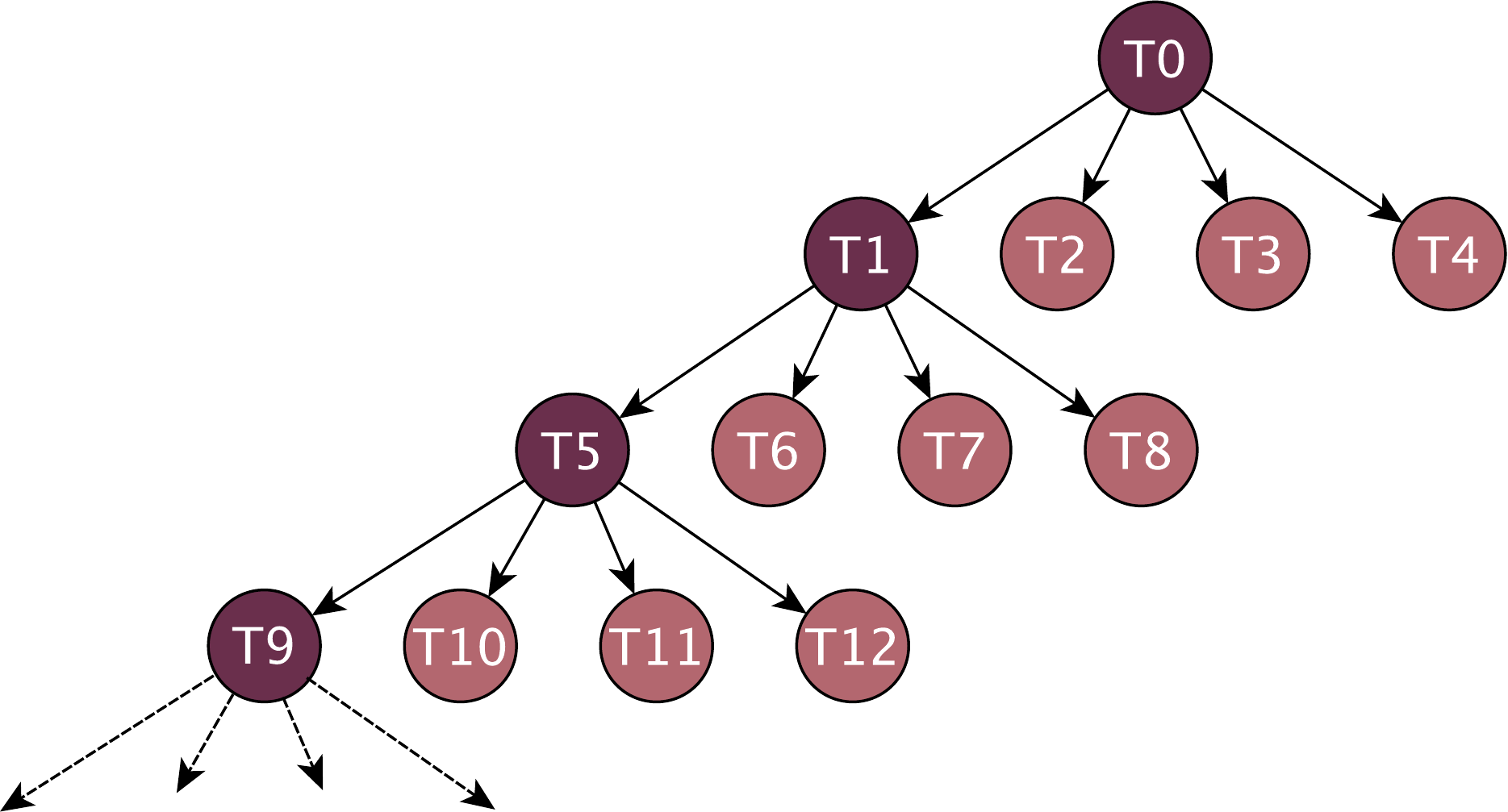}
\caption{An illustration of a task DAG. DAG parallelism is 4. T0, T1, T5 and T9 are marked as high priority tasks.}  \vspace{-2mm}
\label{fig:dag}
\end{figure}

User-specification of task criticality is common in task-parallel models. It supported, for example, by OpenMP task priorities~\cite{openmp45-api}. Criticality can also be inferred dynamically by the runtime system~\cite{chronaki-ics15}. Tasks identified as high priority include tasks that release a large amount of dependent tasks, or tasks that lie on the DAG's critical path. The remaining tasks are then classified as low priority tasks. In the sample DAG shown in Figure~\ref{fig:dag}, tasks T0, T1 and T5, marked with darker color, are high priority tasks while the rest are low priority tasks. 
Another important attribute in this model that describes the DAG's concurrency is the \textit{DAG parallelism}~\cite{blumofe-jacm99}. We define it as the total amount of tasks divided by the length of the longest path. For instance, the partial DAG shown in Figure~\ref{fig:dag}, has a DAG parallelism of 4.
The platform model considers multiple execution resources hereafter referred to both as cores or threads. Cores share the same ISA, but their performance is not necessarily aligned. Per-core performance is determined by static factors such as core asymmetry (big cores, little cores), and dynamic factors such as dynamic voltage-frequency scaling or time-sharing with other processes.  
\testcomment{At runtime, tasks that do not have any unfulfilled dependencies, i.e.~so-called \textit{ready tasks}, are released to worker queues and picked up by cores for execution. Completed tasks may release dependent tasks to the local worker queues once they become ready. Our model assumes an underlying scheduler based on work stealing~\cite{blumofe-jacm99}. Worker threads dequeue and execute tasks from their own work queue before attempting to steal from queues belonging to others.}
We consider that task execution is \textit{moldable}, i.e.~a single task can run in parallel on a variable amount of cores. The set of resources allocated for the execution of a task is called its \textit{execution place}. 
Formally, an execution place is a tuple of two numbers (\textit{core}, \textit{resource width}) where \textit{core} identifies the starting thread number, and \textit{resource width} describes how many threads cooperate to execute the task. Finally, a \textit{resource partition} comprises sets of execution resources. Meaningful resource partitions are those cores that share, for example, cache levels, memory channels, NUMA nodes, etc. 
For instance, NVIDIA Jetson TX2 platform used in our experiments consists of two resource partitions: a dual-core NVIDIA Denver cluster and a quad-core ARM A57 cluster, each with its own shared L2 cache. In this case, the supported resource widths for a task running on the Denver partition are 1 and 2, whereas the supported resource widths for a task on the A57 partition are 1, 2 and 4, as shown in Figure~\ref{fig_resourcewidth}.


\section{ \scheduler}
\label{approach}
\deleted[id=MP]{Traditional} Work stealing schedulers randomly map tasks irrespective of the capabilities of the underlying resources or changes in the execution environment, and can suffer from resource contention and from load imbalance especially when scheduling applications with low DAG parallelism. 
In this section, we present our proposed scheduling solution that adapts to 
various forms of dynamic asymmetry while making no prior assumptions about the underlying architecture or nature of the workload. 

\subsection{Overview}
To mitigate the effects of variability, the scheduler should detect and react to dynamic asymmetry as soon as it appears\deleted[id=MP]{, and react to it by adapting the schedule}. The proposed dynamic asymmetric scheduler builds on two key ideas\deleted[id=MA]{for accomplishing this}. Firstly, it detects dynamic asymmetry through online performance (i.e.~execution time) monitoring of tasks. This leverages the observation that applications typically experience the effects of dynamic asymmetry in the form of performance variation over a period of time.  
Secondly, it schedules to minimize the impact of interference. Two techniques are explored to achieve this: 1) predicting the best possible execution place for high priority tasks according to the online performance model, and 2) molding tasks (i.e.~number of assigned cores) 
to reduce inter-task contention and resource oversubscription \added[id=MA]{, i.e. when a certain hardware resource (e.g. cache level size) does not fit the task requirements}. These techniques allow the scheduler to enhance resource usage and improve throughput by exploiting the DAG parallelism. 
To facilitate detection and adaptation, the scheduler requires an online trace model that predicts the best execution place for a task across the different resource partitions and also if there is potential benefit by enabling moldable execution. We describe the model utilized in this work in Section~\ref{ptt_impl}.

\subsection{Scheduling Algorithm}\label{alg_desc}

\begin{algorithm}[!t] \small
\caption{\textbf{\scheduler}}\label{interfer_algo}
\begin{algorithmic}[1]
\State \textbf{Input:} task type, core id, task priority, Trace Model (TM)
\State \textbf{Output:} execution place
\If{low priority task}
    \State \textit{Local search} to minimize  TM(core id, width)$\times$width
\EndIf
\If{high priority task}
    \If{scheduler == DAM-C}
    \State \textit{Global search} to minimize TM(core id, width)$\times$width
    \EndIf
    \If{scheduler == DAM-P}
    \State \textit{Global search} to minimize TM(core id, width)
    \EndIf
\EndIf
\end{algorithmic}
\end{algorithm}

In this section, we describe the proposed scheduling techniques by introducing our scheduling algorithm targeting dynamic asymmetry. Algorithm~\ref{interfer_algo} describes how ready tasks are assigned their execution places. This algorithm is invoked by the worker after dequeuing a ready task from the local work queue but prior to execution.
In the case of a low priority task, the scheduler attempts to determine the best resource width while keeping the mapping of the task to its local resource partition and the core fixed. This policy enhances data-reuse across dependent tasks. In order to determine the best resource width, the scheduler leverages the trace model \deleted[id=MA] {to obtain execution time prediction for possible resource widths.} \added[id=MA]{, which returns the predicted execution time on a given resource partition. This model is maintained for each task type}. Finally, 
the width that minimizes the parallel cost is selected by minimizing the product of resource width and predicted execution time. We refer to the process of determining the best resource width as \textit{local search} since it involves keeping the resource partition and the core fixed while molding only the resource width.
In the case of a high priority task, the scheduler attempts to determine the best execution place for the task among the different resource partitions in the system. As discussed previously, the trace model is leveraged to obtain execution time prediction for the possible places that a task can be mapped to. The execution place that minimizes parallel cost by finding the lowest product of resource width and predicted execution time is selected. We refer to this process as \textit{global search} since it involves sweeping through all possible execution places.
Given that the number of high priority tasks is usually only a small fraction,
this strategy should, in principle, not result in overcommitting the fast cores. We refer to this scheduler as \textbf{DAM-C} (\textbf{D}ynamic \textbf{A}symmetry scheduler with \textbf{M}oldability, targeting parallel \textbf{C}ost) hereafter since the scheduler strives 
to reduce parallel cost and minimize resource usage. 

In scenarios where parallelism is limited, reducing the parallel cost of tasks can be ineffective as it can lead to increased core idleness. Thus, we propose a variant that performs a global search for critical tasks and selects the execution place that minimizes the predicted execution time. This scheduler is hereafter referred to as \textbf{DAM-P} (\textbf{D}ynamic \textbf{A}symmetry scheduler with \textbf{M}oldability, with critical tasks targeting best parallel \textbf{P}erformance) since the scheduler strives to improve parallel performance of the critical tasks.

\section{Implementation and Experimental Methodology}
\label{setup}
We describe \replaced[id=MA]{an}{the} implementation of the \added[id=MA]{the trace model referred to in Section~\ref{alg_desc}. This implementation is an} online performance model (called PTT) \added[id=MA]{ and is described in} Section~\ref{ptt_impl}\replaced[id=MA]{. In addition,}{and of} the \added[id=MA]{implementation of the} dynamic asymmetry scheduler \added[id=MA]{is described} in Section~\ref{impl_sched}. Both components are implemented on top of XiTAO\footnote{https://github.com/mpericas/xitao.git}, a DAG runtime implemented on top of C++11 designed to evaluate scheduling policies~\cite{pericas-taco18}. Section~\ref{methodology} describes the experimental methodology used to evaluate the schedulers.

\subsection{Implementation Details}
\label{impl}
\subsubsection{Performance Trace Table (PTT)}
\label{ptt_impl}
\begin{figure}[!t]
\centering
\subfigure[Resource Width]{ \includegraphics[width=0.45\columnwidth]{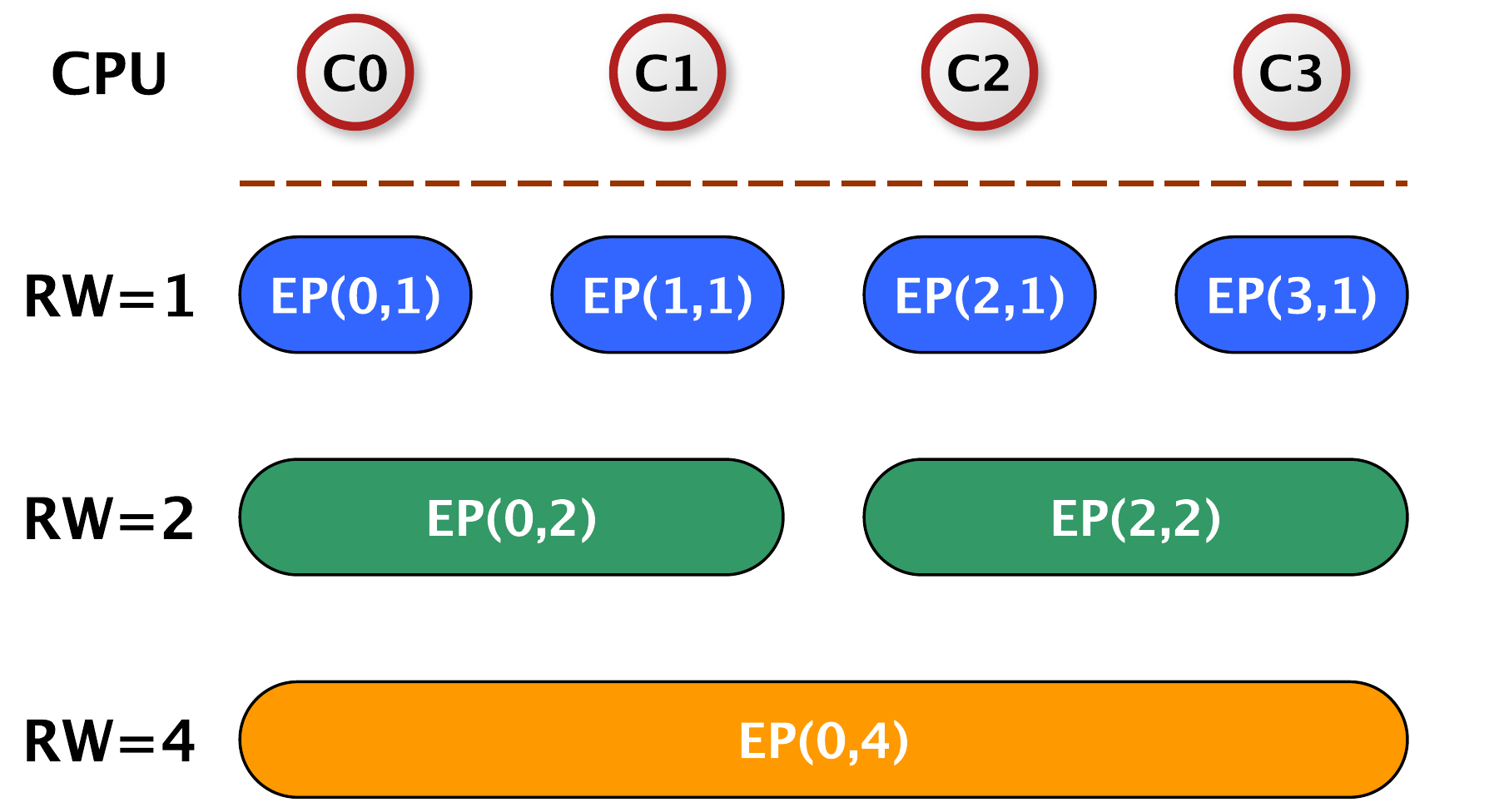}
\label{fig_resourcewidth}}
\subfigure[Performance Trace Table]{
\includegraphics[width=0.45\columnwidth]{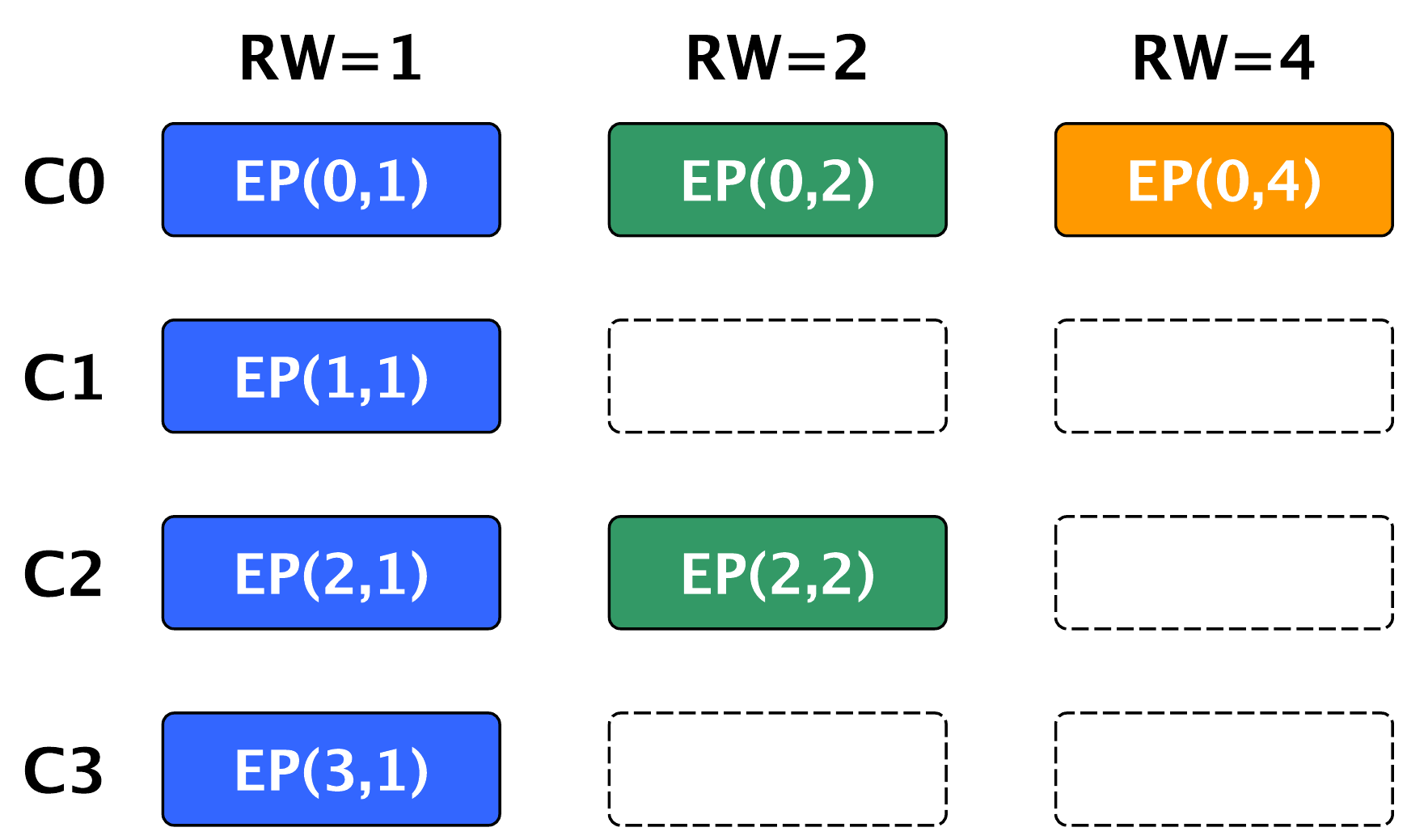}
\label{fig_ptt}}
\caption{PTT organization for four cores. Cx denotes the core number, EP(x,y) represents the task's execution place.} \vspace{-2mm}
\label{fig:time-table}
\end{figure}

Figure~\ref{fig_resourcewidth} shows a representation of a core-cluster with four cores and three possible resource widths of 1, 2 and 4. The corresponding PTT organization is shown in Figure~\ref{fig_ptt}. The goal of this table is to produce a performance estimate for every possible resource partition that can be assigned to a task. 
The number of entries in the table is thus a product of the $\mathit{Number \ of \ cores}$  and the valid $\mathit{Resource \ width}$s.
The entries are initialized to zero. This ensures that all possible execution places are evaluated at least once during the first phases of execution.
Due to the decentralized implementation of the scheduler, the table is organized such that individual rows fit into cache lines.  Each core mainly accesses a single cache line indexed with its own core id, keeping access latency low.  
Each entry of the PTT keeps track of the execution time of the task, as observed by the leader core, with a specific width, which significantly simplifies the implementation.
To avoid large fluctuations in the values of the PTT, which could result from short isolated events, 
we update the PTT entries by computing a weighted average.

Our sensitivity analysis (see Section~\ref{Perf_SA}) suggests that each entry be updated with a weighted ratio of 1:4, that is, 
$updated\_value=[(4 \times old\_value) + 1 \times new\_value)]/5$. This averaging ensures that after a performance variation, at least three measurements need to be taken before the PTT value becomes closer to the new value. 
Although weighted averaging results in an additional PTT read operation, it is crucial to be resilient to divergent measurements as the PTT is critical for making scheduling decisions. 
One such table is instantiated for each task type as the performance varies per type. 

Setting up the PTT 
only requires information about the number of cores and their organization into core-clusters with shared caches. This information can readily be obtained using tools like $hwloc$~\cite{broquedis-hwloc-pdp10}.
Upon completion of \replaced[id=MP]{each}{a} task, the workers simply update the corresponding index in the PTT 
thus implementing a dynamic online model for performance prediction for a particular task type. \added[id=MP]{A task type refers to each function implemented as a task. Within XiTAO it refers to the C++ class describing the functionality. Note that there is one PTT for each task type. Tasks execution times are measured during normal execution, instead of a profiling phase. The measured execution times are thus impacted by co-scheduling of other tasks in the system and the shared resource interference they generate. As long as the application behavior does not change quickly, estimates generated by the performance model will be aware of co-scheduled tasks.}
Although performance prediction using PTT is simple, our evaluation, in Section~\ref{Perf_Eva} using different platforms, shows that this approach is effective for improving performance in the presence of dynamic asymmetry. 
\added[id=JC]{The overhead of globally searching the whole PTT is in the order of one microsecond in our evaluation platform (NVIDIA Jetson TX2). We are aware that the design, however, may result in non negligible overheads when scaling to platforms with large amount of execution places and cores. 
The design and evaluation of scalable performance prediction models is left for future work.}

\subsubsection{\scheduler}\label{impl_sched}
\begin{figure}[t]
\centering
\includegraphics[width=1\columnwidth]{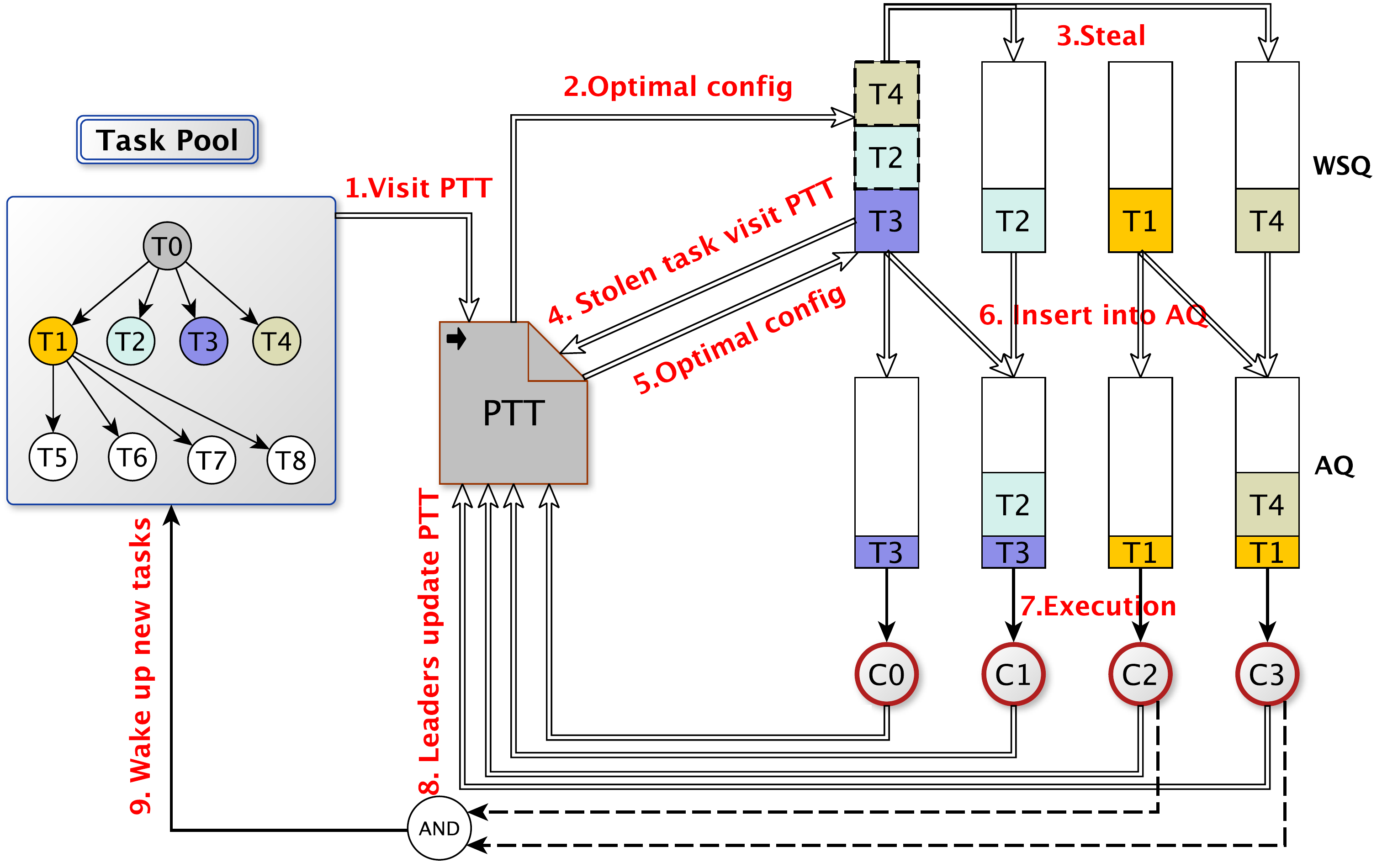}
\caption{Implementation overview of \scheduler\/ on top of the XiTAO runtime system.}  \vspace{-1mm}
\label{fig:perf_sched}
\end{figure}

The scheduler is implemented as an extension on top of the XiTAO runtime system. 
XiTAO relies on work stealing~\cite{blumofe-jacm99} for assigning tasks to workers and supports moldable execution of tasks.  
This is achieved by implementing two queues for each worker: a Work Stealing Queue (WSQ) and a FIFO Assembly Queue (AQ)~\cite{pericas-taco18}.
The WSQs hold the ready tasks and use random work stealing for load balancing.
The actual execution place is selected only after a task becomes ready (i.e. after all its input dependencies have been satisfied). 
At this point, pointers to the tasks are inserted into all AQs representing the execution place for the task, from where they are finally executed by the cores. 
We disable the stealing of high priority tasks in order to guarantee that all such tasks are executed according to their scheduling decision (cores id and optimal resource width). 
Low-priority tasks, on the other hand, are subject to random work stealing by idle workers. 
Figure~\ref{fig:perf_sched} illustrates the operation of the scheduler when executing the task DAG shown in Figure~\ref{fig:dag}.
The steps denote the lifetime of a task from wake-up (release by predecessor) to commit (finalization).
We assume that T0 has been executed on core 0 with resource width=1 and has updated the PTT entry (c=0,w=1). Core 0 then wakes up the child tasks of T0, namely T1 (high priority task), T2, T3 and T4 (low-priority tasks).
During step 1 and 2 (shown in Figure~\ref{fig:perf_sched}), ready tasks read the PTT to determine the best execution place. 
For this example, we assume that the best PTT configurations for  T1, T2, T3 and T4 are (2,2), (0,2), (0,2) and (0,2), respectively.
For T1, we globally search the PTT to determine the best execution place and insert the task in the WSQ of core 2, 
while for the other tasks, i.e. T2, T3, T4, 
we insert them into the local queue of their parent T0, i.e.~the WSQ of core 0.
When the WSQs of cores (C1,C3) are empty,
the workers attempt to steal low-priority tasks from other WSQs that have more tasks (as indicated in step 3).
 Cores 1 and 3 successfully manage to steal T2 and T4 from the WSQ of core 0.
After a successful steal, the PTT is visited again (step 4) to determine the best execution place.
For T2 and T4, width=1 is chosen after performing a local search of the PTT again (instead of width=2 from the earlier search) as indicated in step 5.
After this step, these tasks are distributed to the corresponding AQs as indicated in step 6.
In step 7, the cores fetch task partitions from their own AQs for execution.
After the leader core finishes execution, it updates the corresponding PTT entries using the weighted update described in Section~\ref{ptt_impl}, which is necessary for training the table (step 8).
As cores 2 and 3 can execute T1 asynchronously, the runtime takes no further action until both cores finish executing the priority task.
The last core to complete the execution of the task then wakes up the dependent tasks from the task pool (T5 to T8).

\subsection{Experimental Methodology}
\label{methodology}
\subsubsection{Evaluation Platforms}
We evaluate the proposed scheduler on two platforms. The first is an NVIDIA Jetson TX2 development board, featuring a dual-core NVIDIA Denver 2 64-bit CPU and a quad-core ARM A57 cluster (each with 2 MB L2 cache). The platform is asymmetric since the Denver cores are generally faster than the A57 cores. We use this platform to evaluate interference due to co-running applications and scenarios of DVFS interference on fixed asymmetric platform. The code is compiled using \texttt{gcc 5.4.0} on \texttt{Linux version 4.4.38-tegra}.
The second platform consists of four dual-socket 10-core Intel 2650v3 (code-named "Haswell") nodes connected via Mellanox ConnectX-3 FDR Infiniband and run \texttt{Linux version 3.10.0}. We use the Haswell cores as a symmetric platform to evaluate interference due to co-running applications, compiled using the \texttt{icpc (ICC) 19.0.5.281} compiler. For MPI code, we leverage the \texttt{Intel MPI Library v2018}. 

\subsubsection{Benchmarks}
We evaluate the described scheduling techniques using synthetic benchmarks and \added[id=JC]{two} applications \added[id=JC]{: K-means and Heat}.
\deleted[id=JC]{The purpose of the synthetic benchmark is to test various types of kernels and parallelism. We first describe the synthetic benchmark, followed by three applications: VGG-16, K-means and Heat.}

\textbf{Synthetic Directed Acyclic Graph:}
We construct synthetic DAGs with the notion of priority tasks and with specific DAG parallelism. In the DAG, each layer consists of a same number of tasks $P$, equal to the DAG parallelism, and same type of task. One of the tasks is marked as critical. Upon the execution of the critical task, another set of $P$ tasks with the same characteristics are released. The constructed DAG comprises nodes that represent one of three kernels: Matrix Multiplication, Copy and Stencil. Each DAG node (task) is moldable and can execute on a variable number of processors. The node types are described below: 

\textbf{Matrix Multiplication} 
\deleted[id=JC]{A \textit{matrix multiplication} DAG benchmark} represents the \textit{compute-intensive} class of workloads. Matrices $A$, $B$ and $C$ are pre-allocated and partitioned in tiles of $N \times N$. 
This kernel executes general matrix multiplication calls within the graph nodes.

\textbf{Copy} 
\deleted[id=JC]{A \textit{copy} DAG benchmark} represents the \textit{memory-intensive} class of workloads.
This kernel reads and writes large portions of data to memory, effectively creating a streaming behavior whereby the main memory is accessed continuously. 

\textbf{Stencil}
\deleted[id=JC]{A \textit{stencil} kernel}
represents the \textit{cache-intensive} class of workloads. 
It performs repeated updates of values associated with points on a multi-dimensional grid using the values at a set of neighboring points. 
 
Unless specified otherwise, the matrix-tile-size (per task) for the \textit{MatMul} kernel is 64$\times$64, whereas the number of tasks in the DAG is 32000.
In the case of \textit{Copy} and \textit{Stencil} kernel the matrix tile-size (per task) is 1024$\times$1024, and number of tasks in the DAG is 10000 and 20000, respectively. 
\testcomment{Furthermore, when evaluating the different classes of synthetic DAGs, we assume sweep through DAG parallelism from 2 to 6. This is primarily because the evaluation platform has 6 cores. 
Additionally, low parallelism DAGs are not uncommon in real applications, which are often limited by synchronization and inter-task data dependencies. This is also in line with a recent study~\cite{feng-ispass19} that found that the average thread parallelism among modern desktop applications is low ($\approx 3$).}

\deleted[id=JC]{
\paragraph{Image Classification (Darknet-VGG-16 CNN):} \label{vgg_setup}
The first application is a port of the well-known VGG image classification network~\cite{simonyan2014very}.
We have ported the VGG-16 CNN (Convolutional Neural Network) implementation from Darknet~\cite{darknet13} to the XiTAO runtime.
VGG-16 is a 16-layer deep neural network. The first 13 layers are convolution layers, while the last 3 layers are fully connected layers. The XiTAO VGG-16 is implemented as a fork-join DAG that spans all the layers. Each node of the DAG performs  GEneral Matrix Multiply (GEMM) operations on a sub-partition. 
}

\begin{figure*}[!t]
\centering
 \subfigure[Matrix Multiplication]{\label{fig_wi_mm_BC}  \includegraphics[width=0.32\textwidth]{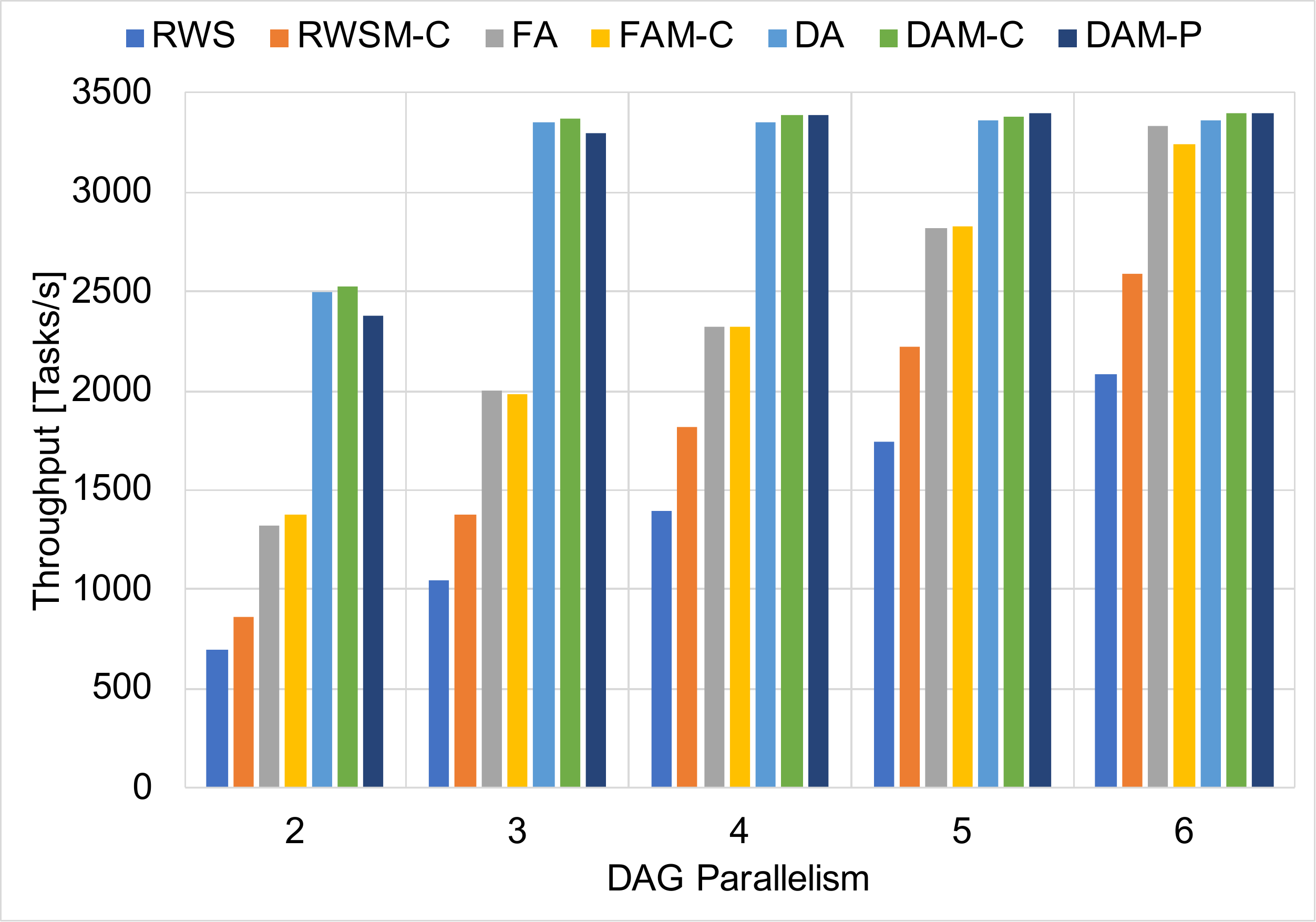}}
 \subfigure[Copy]{\label{fig_wi_copy_BC} \includegraphics[width=0.32\textwidth]{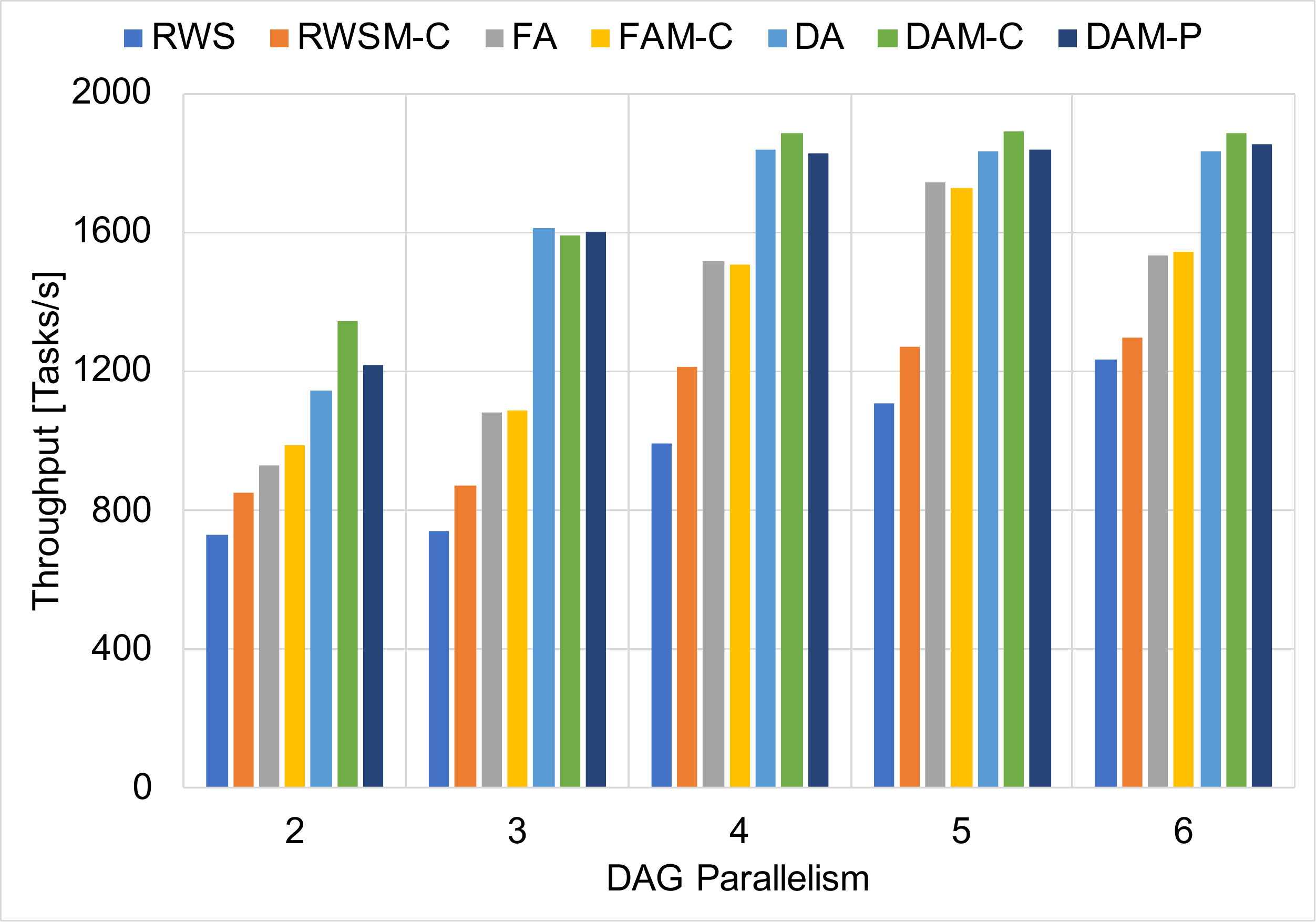}}
 \subfigure[Stencil]{\label{fig_wi_stencil_BC} \includegraphics[width=0.32\textwidth]{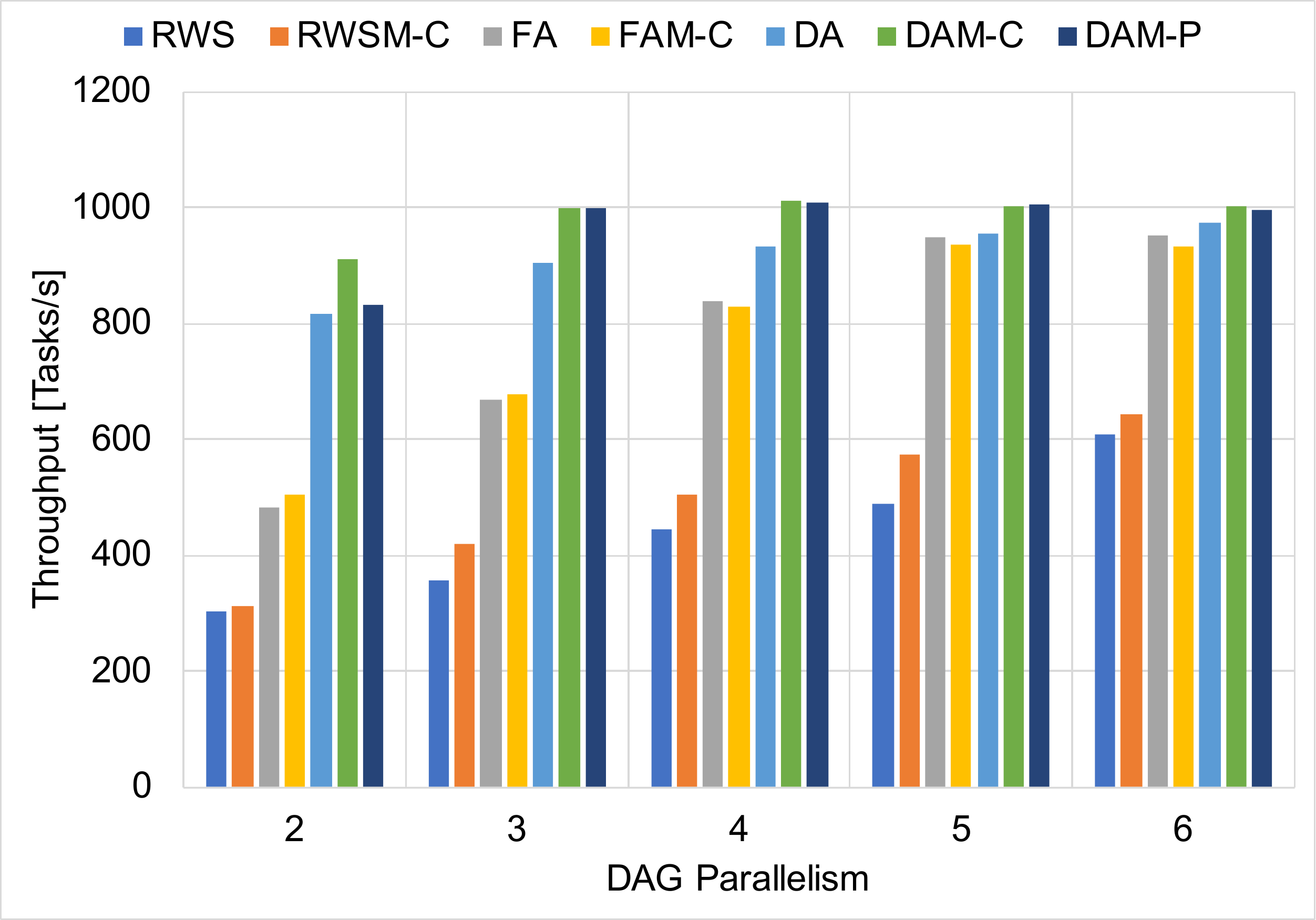}}
 \caption{The performance impact of co-running application interference and comparison between different schedulers with DAG parallelism ranging from 2 to 6.} 
 \label{with_interference_BC}
\end{figure*}

\textbf{K-means Clustering} \added[id=JC]{application is selected}
\deleted[id=JC]{We use the K-means clustering} from the Rodinia Benchmark suite~\cite{rodinia}. 
\replaced[id=JC]{It}{This benchmark} is a  representative of the data-parallel class of applications. The XiTAO runtime interface supports loop-parallel constructs, and provides the ability to tune the granularity of the loop task partitions and to nest the loop in a graph node. These features are utilized in our implementation to describe K-means as a dynamic DAG.

\textbf{Distributed 2D Heat}
\replaced[id=JC]{is implemented as}{We implement} an iterative distributed 2D stencil in which MPI calls are encapsulated into specific TAOs that are responsible for exchanging the boundaries (ghost cells). There is one such exchange per iteration. Due to the criticality of such communication, these MPI tasks are marked as high priority tasks.

\subsubsection{Scheduling Policies}
To evaluate the various scheduling techniques described in this paper in the context of both fixed and dynamic performance asymmetry, we evaluate a set of scheduler configurations on the TX2 platform\deleted[id=MM]{, ranging from simple random work stealing policies to the two proposed policies: DAM-C and DAM-P}. \replaced[id=MM]{The}{All} scheduling techniques considered during the evaluation are summarized in Table~\ref{table:schedulers}.
The \deleted[id=MM]{most basic scheduler is a} random work-stealing scheduler (RWS), \deleted[id=MM]{which} behaves as a decentralized greedy scheduler where each thread owns a work-stealing queue.  
Irrespective of their priority, child tasks are pushed to the local queues and allowed to be stolen in order to mitigate load-imbalance. 
We then extend this scheduler with the option of moldability (RWSM-C), which allows to aggregate resources and execute tasks over a set of cores with the goal of improving parallel cost in a similar fashion to DAM-C. The implementation of RWSM-C requires the integration of a performance model to select a number of cores. To support this option, RWSM-C includes an implementation of the performance trace table (PTT), as described in Section~\ref{ptt_impl}. 
The third scheduler is a criticality-based scheduler designed for fixed (static) asymmetric architectures, called Fixed Asymmetry Scheduler (FA). 
In this scheme, priority information is processed by the runtime and high-priority tasks are strictly mapped to statically faster cores (the Denver cores in the context of the TX2 platform). The scheduler is inspired by prior works like the Critical-Path-on-a-Processor algorithm~\cite{topcuoglu-tpds02} and the Criticality-Aware Dynamic Task-Scheduling (CATS)~\cite{chronaki-ics15}, both of which are based on the assumption that performance asymmetry remains unchanged over time.
Unlike CATS, our work does not address the problem of determining task criticality dynamically. Hence, FA and FAM-C (described next) rely on the static scheme described in Section~\ref{background}. 
Similar to RWS, we extend the FA scheduler with the option of moldability targeting parallel cost. This yields the fourth configuration called FA+Moldability (FAM-C). 
Finally, we also implement a variant of the DAM schedulers but without moldability, simply called Dynamic Asymmetry Scheduler (DA). 
This scheduler only searches for the fastest core on which to execute critical tasks and then executes these tasks on a single core.  
All these different scheduler configurations help us isolate and evaluate the additional gains introduced by incorporating dynamic asymmetry awareness, criticality-aware scheduling and task moldability. We evaluate their impact in the next section. 

\begin{table}[h]\scriptsize
\caption{Features summary of all evaluated schedulers}  \vspace{-2mm}
\begin{tabular}{|c|c|c|c|}
\hline
\textbf{Name} & \textbf{[A]symmetry awareness} & \textbf{[M]oldability} & \textbf{Priority placement} \\ \hline
 RWS
 & N/A       & N/A       & N/A                  \\ \hline
 RWSM-C                  & N/A     & Yes    & Resource [C]ost             \\ \hline
 FA
 & [F]ixed   & No        & N/A                  \\ \hline
 FAM-C                     & [F]ixed   & Yes       & Resource [C]ost  \\ \hline
 DA                 & [D]ynamic         & No        & N/A \\ \hline
 DAM-C            & [D]ynamic & Yes       & Resource [C]ost  \\ \hline
 DAM-P             & [D]ynamic & Yes       & [P]erformance          \\ \hline
\end{tabular}
\label{table:schedulers}
\end{table}

\section{Performance Evaluation} \label{Perf_Eva}

The goal of the evaluation is to understand the performance impact of the schedulers depicted in Table~\ref{table:schedulers} during episodes of interference. We evaluate two common scenarios that introduce dynamic asymmetry during application execution. The first is interference due to co-running applications. The second is 
DVFS-based interference due to power management. We evaluate the impact of dynamic asymmetry on an asymmetric as well as on symmetric hardware platforms.
We begin in Section \ref{Perf_IA} by evaluating and analyzing the performance impact of interference arising from co-running applications. Following this, in section \ref{Perf_DVFS}, we study the impact of DVFS, while in Section~\ref{Perf_SA} we evaluate the sensitivity of the scheduler parameters to the task granularity and the weighted update strategy of the PTT. These evaluations are all conducted on the asymmetric NVIDIA Jetson TX2 platform.
We then evaluate the impact of co-running applications together with the three applications on the symmetric Intel Haswell platform in Section~\ref{Perf_Haswell}.

\begin{figure*}[t]
\centering
\subfigure[RWS]{\label{fig_pc_homo} \includegraphics[width=0.13\textwidth]{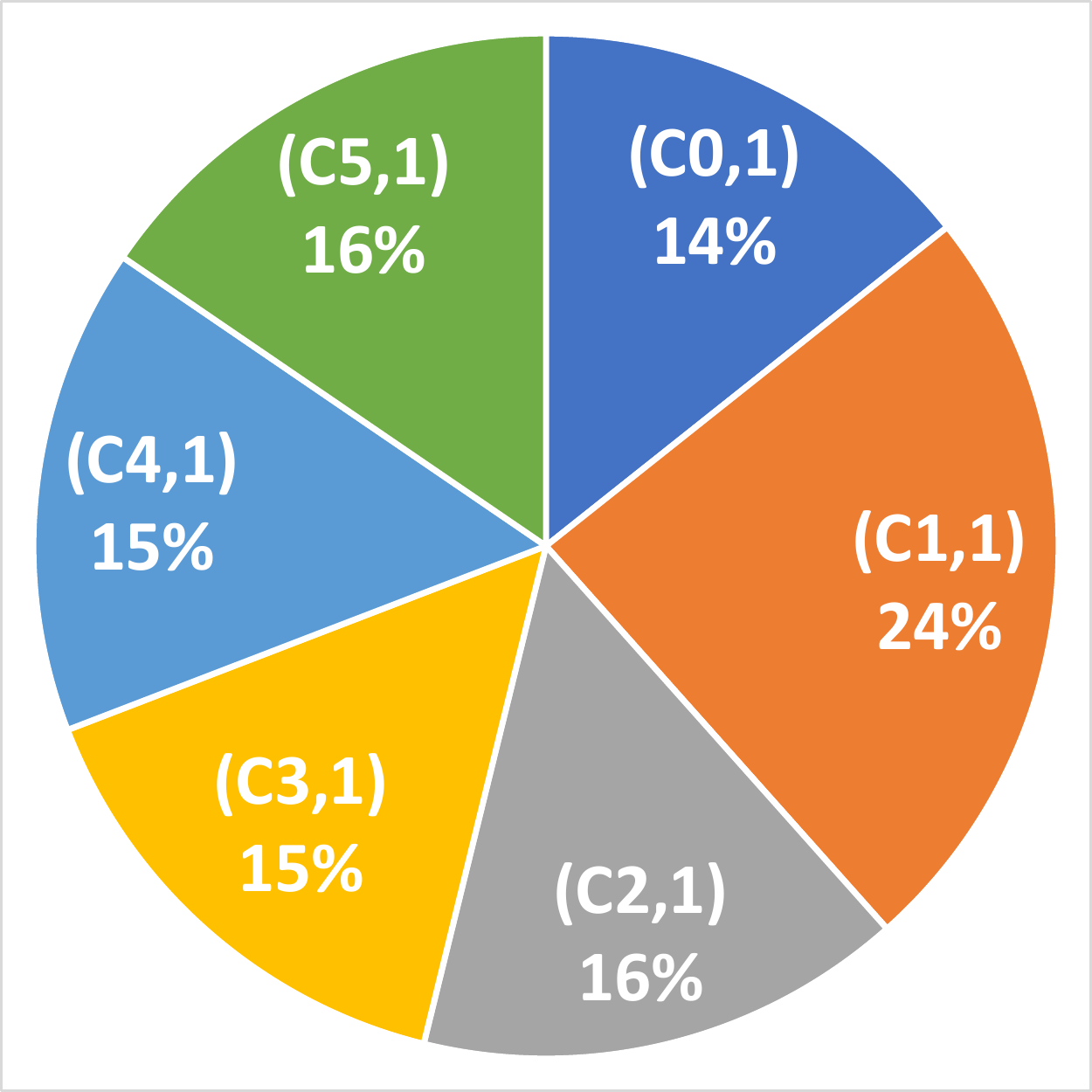}}
\subfigure[RWSM-C]{\label{fig_pc_homo_m} \includegraphics[width=0.13\textwidth]{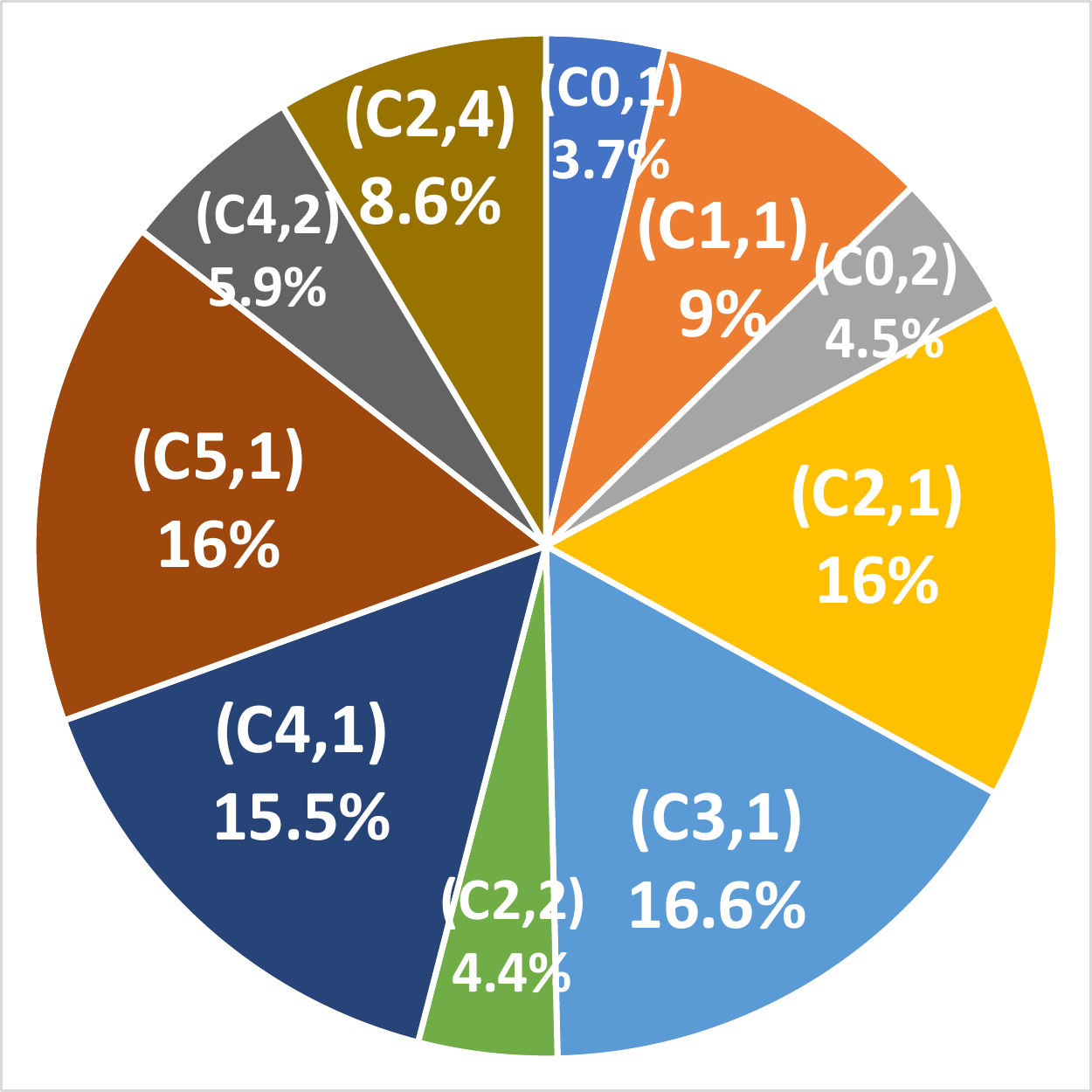}}
\subfigure[FA]{\label{fig_pc_cats}
\includegraphics[width=0.13\textwidth]{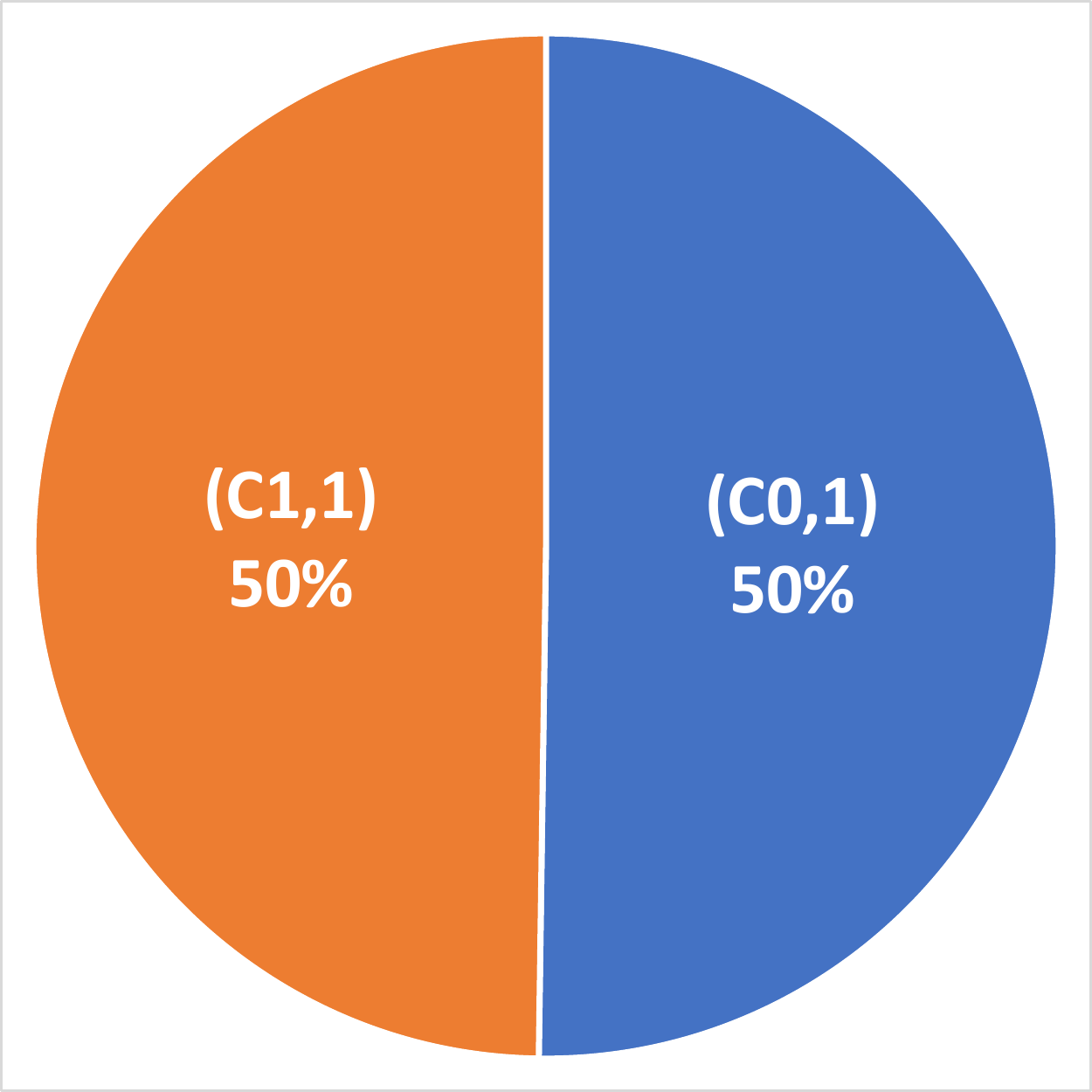}}
\subfigure[FAM-C]{\label{fig_pc_cats_m} \includegraphics[width=0.13\textwidth]{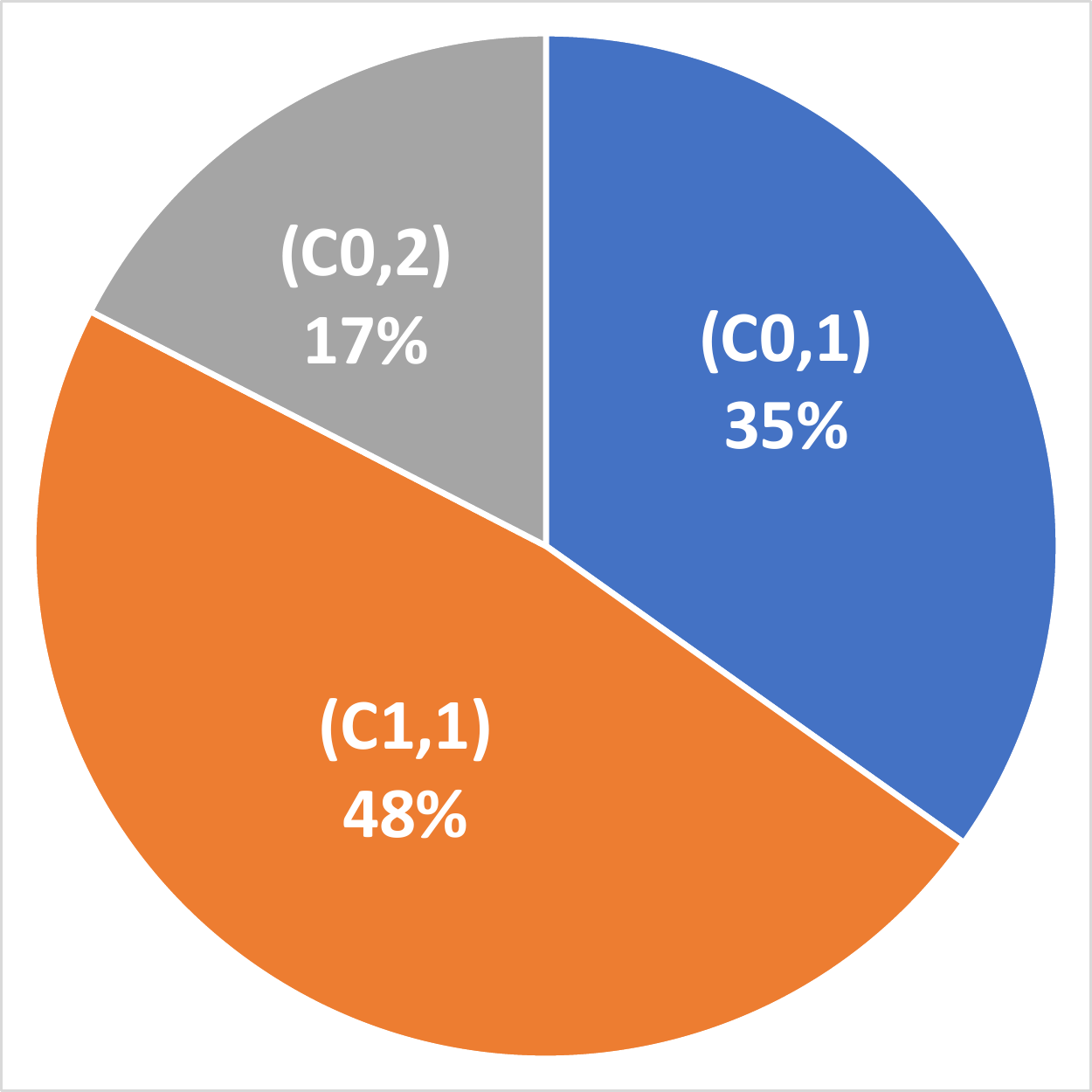}}
\subfigure[DA]{\label{fig_pc_ptt_nm} \includegraphics[width=0.13\textwidth]{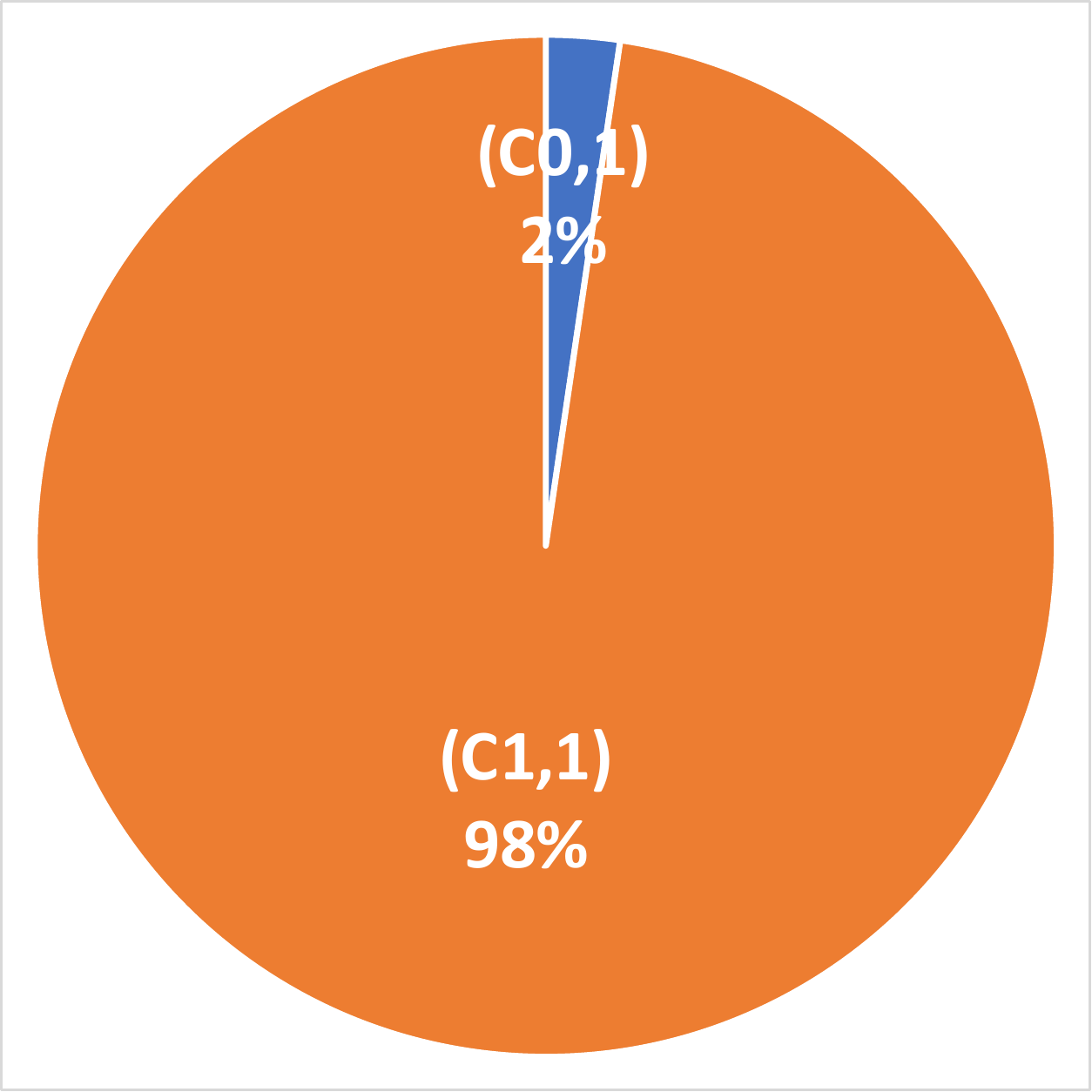}}
\subfigure[DAM-C]{\label{fig_pc_ptt} \includegraphics[width=0.13\textwidth]{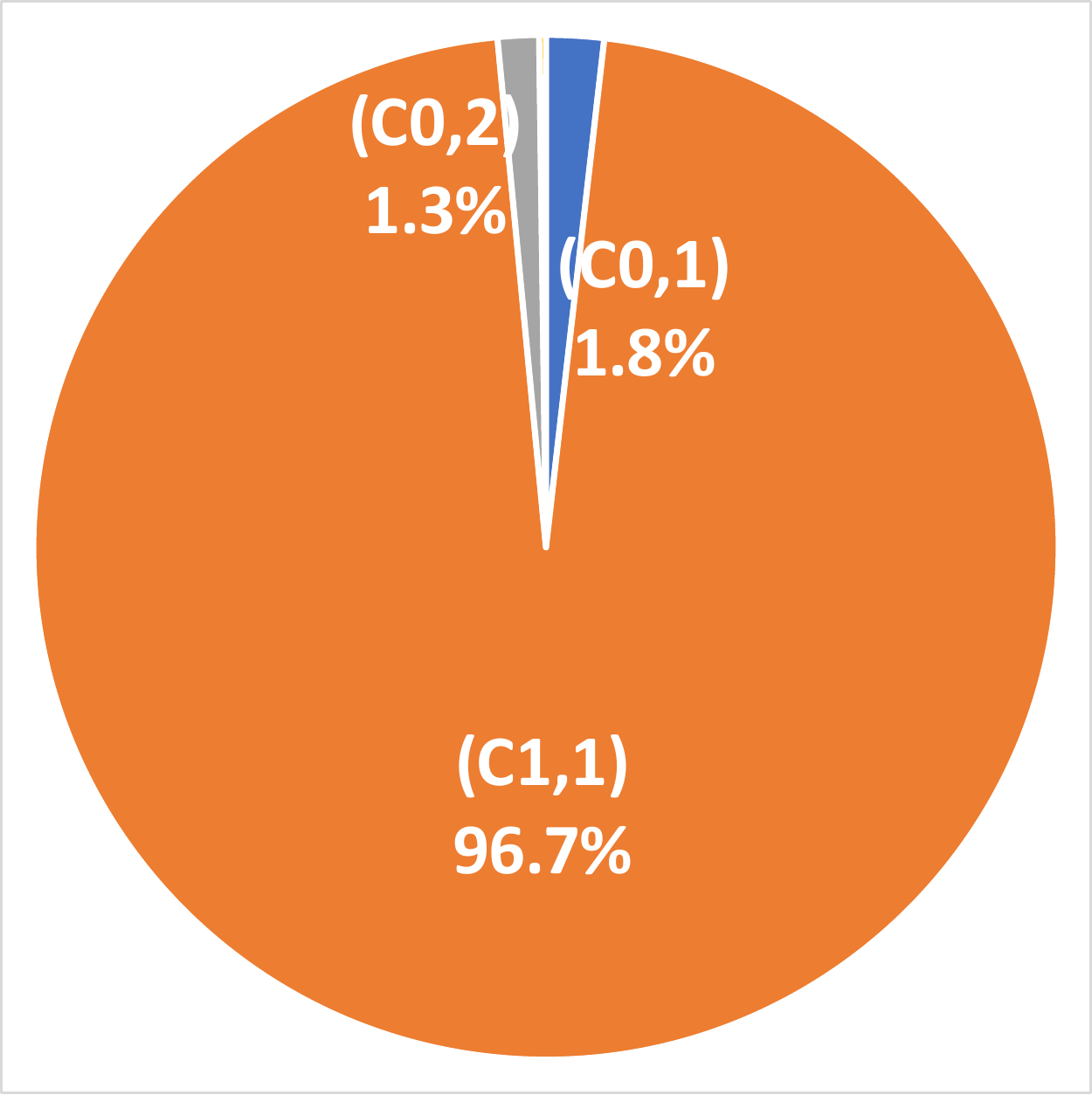}}
\subfigure[DAM-P]{\label{fig_pc_ptt_cnp} \includegraphics[width=0.13\textwidth]{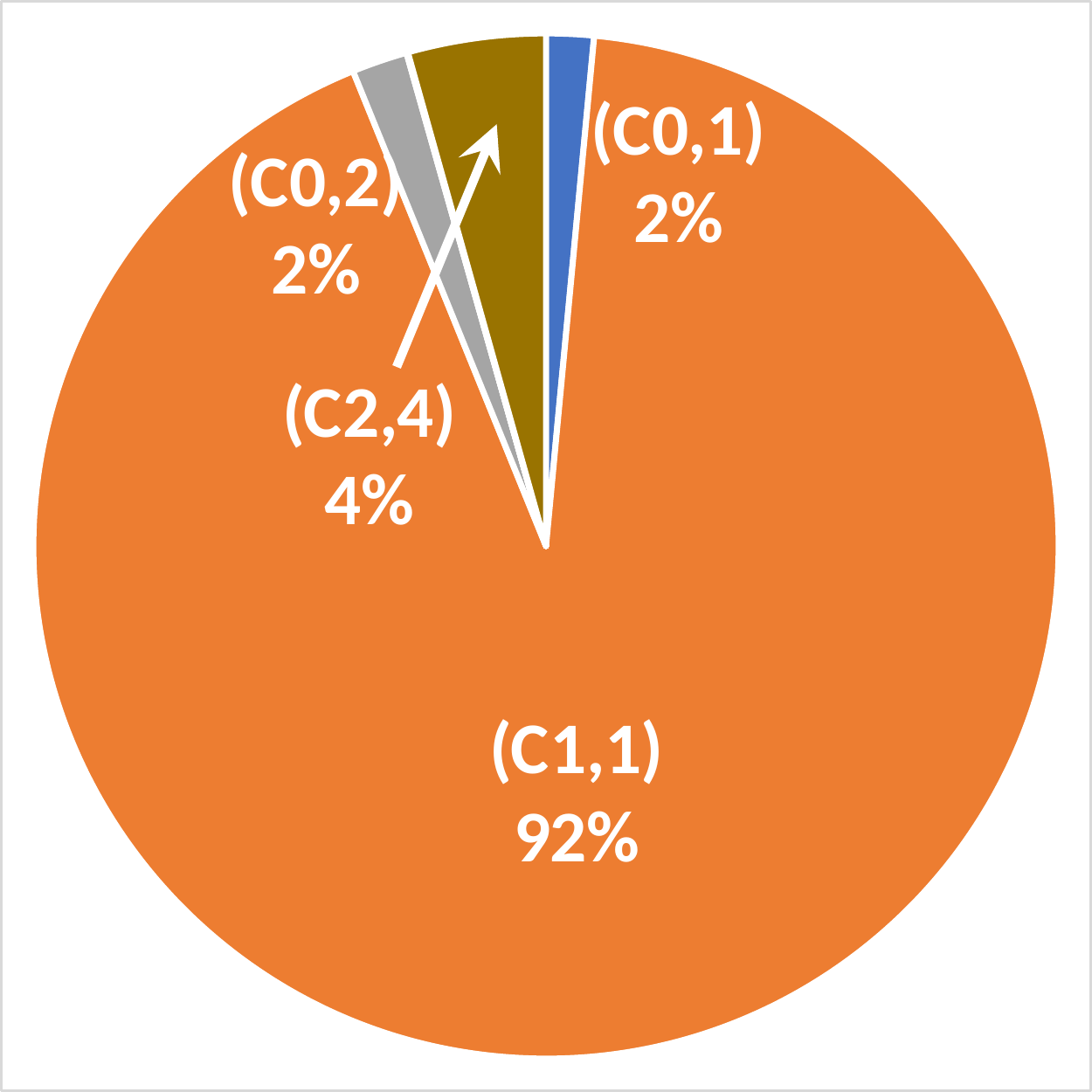}}
\caption{Distribution of priority tasks on each core.} 
\label{fig:numtasks_pc}
\end{figure*}

\subsection{Dynamic Asymmetry Awareness}
\label{Perf_IA}

\begin{figure}[h]
\centering
\includegraphics[width=0.40\textwidth]{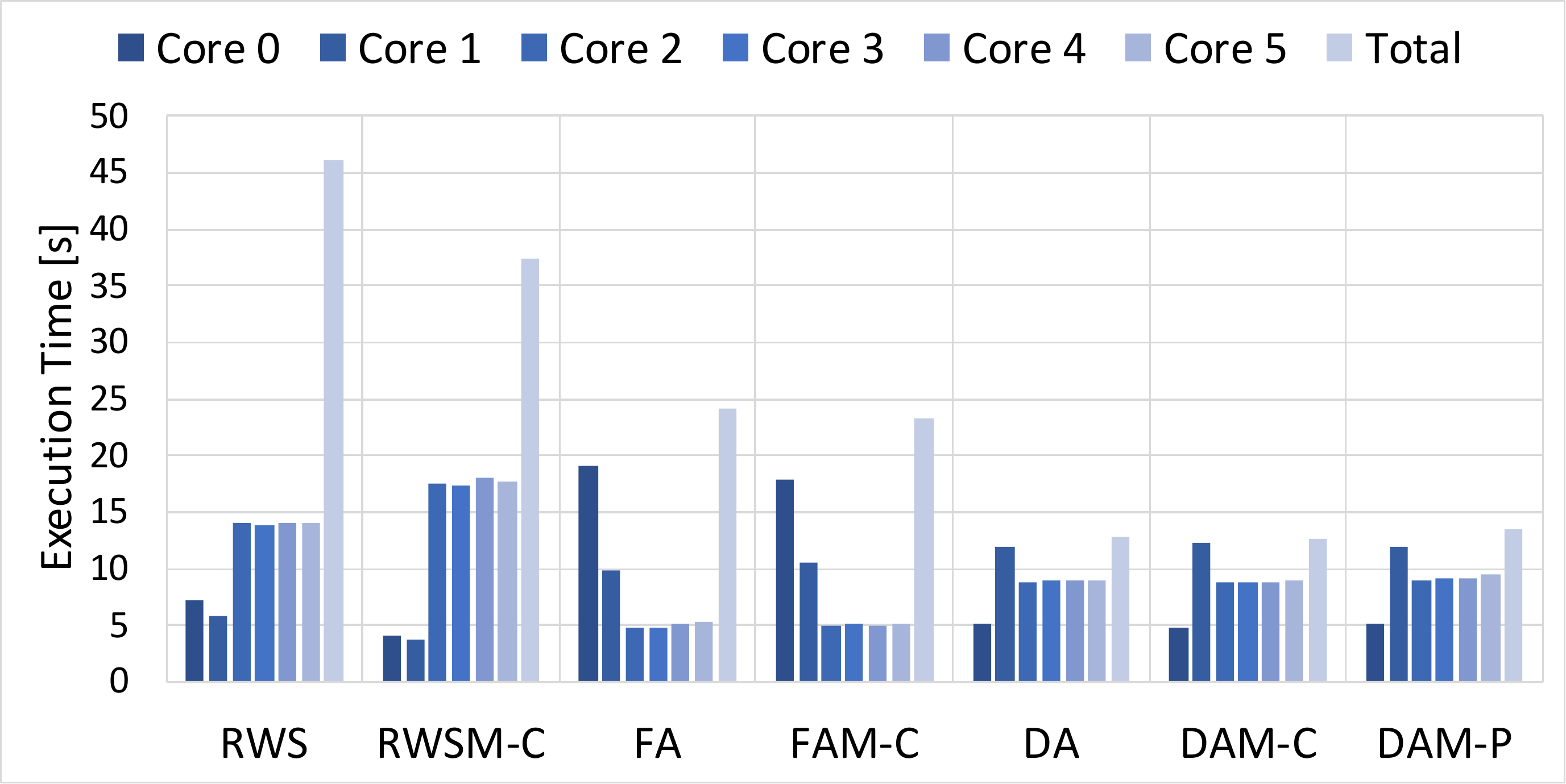}
\caption{Execution time of each thread while co-running application runs on Denver core 0.} 
\label{fig:exec_time}
\end{figure}

Figure~\ref{with_interference_BC} presents the throughput of different schedulers in the presence of a co-running application on core 0 that persists throughout the whole execution time. 
The throughput numbers are computed by dividing the total number of tasks by the total execution time. 
For the case of \textit{matrix multiplication} and \textit{stencil} synthetic DAG, the co-running application consists of a single chain of tasks composed of \textit{matrix multiplication} kernels. This results in CPU interference. 
For the case of the \textit{copy} synthetic DAG, the co-running application is a single task chain of \textit{copy} kernels, which results in memory interference.
The results show that the \scheduler\/s, including DA, DAM-C and DAM-P, provide the highest throughput for the different levels of DAG parallelism and across different kernels, thanks to the ability to adapt to interference. Although the fixed asymmetry schedulers (FA and FAM-C) provide higher throughput than the random work stealing variants by exploiting knowledge of task criticality, they still leave considerable room for improvement. 
For the compute-bound \textit{matrix multiplication} kernel, DAM-C achieves up to 3.5$\times$ speedup compared to RWS.
Also, DAM-C achieves up to 90\% and 85\% performance enhancement compared respectively to FA and FAM-C across different levels of DAG parallelism.
Although DAM-P for \textit{matrix multiplication} is slightly worse than DA and DAM-C on parallelisms 2 and 3, it still achieves much higher throughput than other schedulers.
In Figures~\ref{fig_wi_mm_BC} and~\ref{fig_wi_stencil_BC}, we can observe that the performance of RWS, FA and FAM-C is somewhat linearly proportional to the DAG parallelism, while DAM-C and DAM-P already achieve close to the maximum throughput when parallelism is low.
We conduct additional analysis to understand the gains from dynamically steering the execution of priority tasks using our scheduler in dynamic environments. 
For this, we consider \textit{matrix multiplication} with DAG parallelism of 2 (and in presence of co-running application on core 0).
In this configuration, high priority tasks constitute 50\% of the entire DAG. 
Additionally, cores 0 and 1 refer to the two Denver cores, while core 2 to 5 refer to the four A57 cores.
Figure~\ref{fig:numtasks_pc} presents the distribution of priority tasks on cores for each scheduler.
The labels on the pie chart denote the execution place.
For example, (C2,4) means that the starting core is 2 and the task resource width is 4, i.e. it runs on cores \{2,3,4,5\}.
Figure~\ref{fig:exec_time} shows the cumulative work time for each thread.  
Note that it shows the accumulation of kernels' work time on each core excluding the runtime's activity and idleness.


Figure \ref{fig_pc_homo} shows that with RWS, the priority tasks end up being almost uniformly distributed across the 4 A57 cores (around 15\% each). The faster Denver core ends up executing the largest fraction (24\%) of the high priority tasks while the Denver core that experiences interference executes the least amount (14\%) of high priority tasks. 
Since the background activity is running on Denver core 0, core 1 is subsequently faster, which results in 10\% more total tasks (with a lower worktime as shown in figure~\ref{fig:exec_time}) running on core 1. 
Although the RWS scheduler does not show a dramatic jump in the work time of core 0, it still shows the worst performance as it does not consider task priority.
Figure~\ref{fig_pc_homo_m} shows that only 17.2\% (3.7\%+9\%+4.5\%) of priority tasks are executed on Denver cores because of the interference when the moldability is enabled in RWSM-C, which leads to less task stealing from A57 to Denver cores compared with RWS.
In this case, RWSM-C with interference on Denver core 0 is more load balanced than RWS when the parallelism is low, since more priority tasks complete the execution on A57 cores and can release new tasks faster.
FA has a fixed notion of system asymmetry and strictly assigns the priority tasks to the faster cores, i.e. Denver cores 0 and 1.
Figure~\ref{fig_pc_cats} reflects this fact by showing all priority tasks equally executed on the two Denver cores. 
However, it is obvious that FA is not adaptive to interference, which results in the highest execution time on core 0 as shown in Figure~\ref{fig:exec_time}.
When moldability is enabled (FAM-C), 17\% of the critical tasks execute on both cores 0 and 1 (resource width = 2).
Core 1 runs 13\% more tasks than core 0 due to interference on core 0.
However, FAM-C is still restrictive in the treatment of priority, because it has no means for migration of priority tasks. 
This appears as a loss in throughput compared to DA, DAM-C and DAM-P, due to the delays in releasing the DAG parallelism.
Finally, the dynamic schedulers show more interference awareness than other schedulers and exhibit similar trend across DA, DAM-C and DAM-P as shown in Figure~\ref{fig:numtasks_pc} and ~\ref{fig:exec_time}.
DA only has as few as 2\% of all priority tasks running on the interfered core 0. The majority of tasks are migrated to the faster core of the platform. 
This ensures that both the background application and the foreground DAG are not affected significantly.
When moldability is used (DAM-C), 1.3\% of priority tasks are executed on (C0,2).
DAM-P magnifies the impact of priority by selecting the fastest possible execution place. 
Figure~\ref{fig_pc_ptt_cnp} shows that with DAM-P most of tasks (92\%) execute on the faster Denver core 1 with resource width 1. 4\% of tasks execute on four A57 cores (width = 4), since at some points DAM-P predicts that spanning the whole A57 cluster is faster during interference. 

\subsection{DVFS Awareness}
\label{Perf_DVFS}

\begin{figure*}[t]
\centering
\subfigure[MatMul]{\label{fig_dvfs_mm} \includegraphics[width=0.32\textwidth]{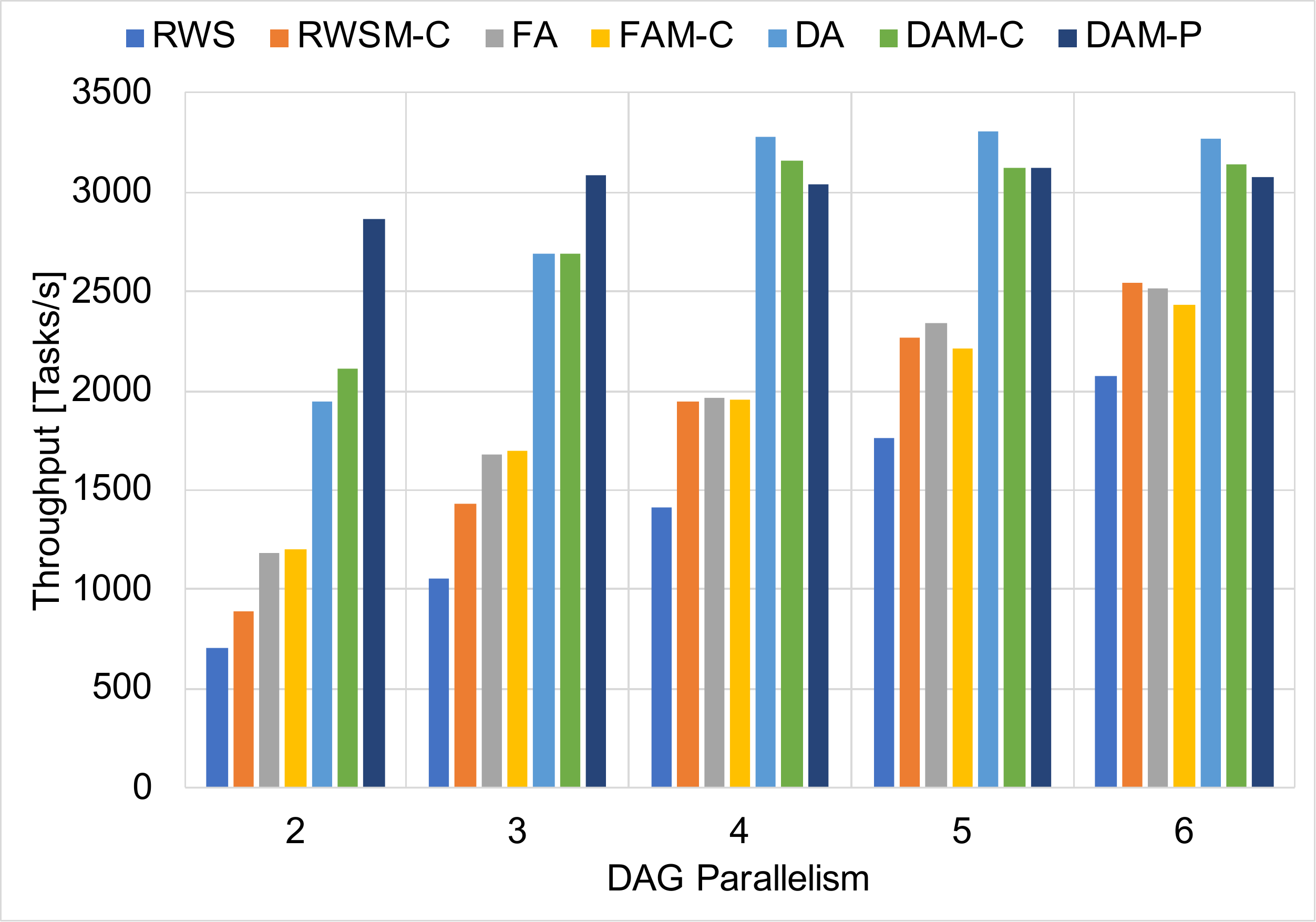}}
\subfigure[Copy]{\label{fig_dvfs_copy} \includegraphics[width=0.32\textwidth]{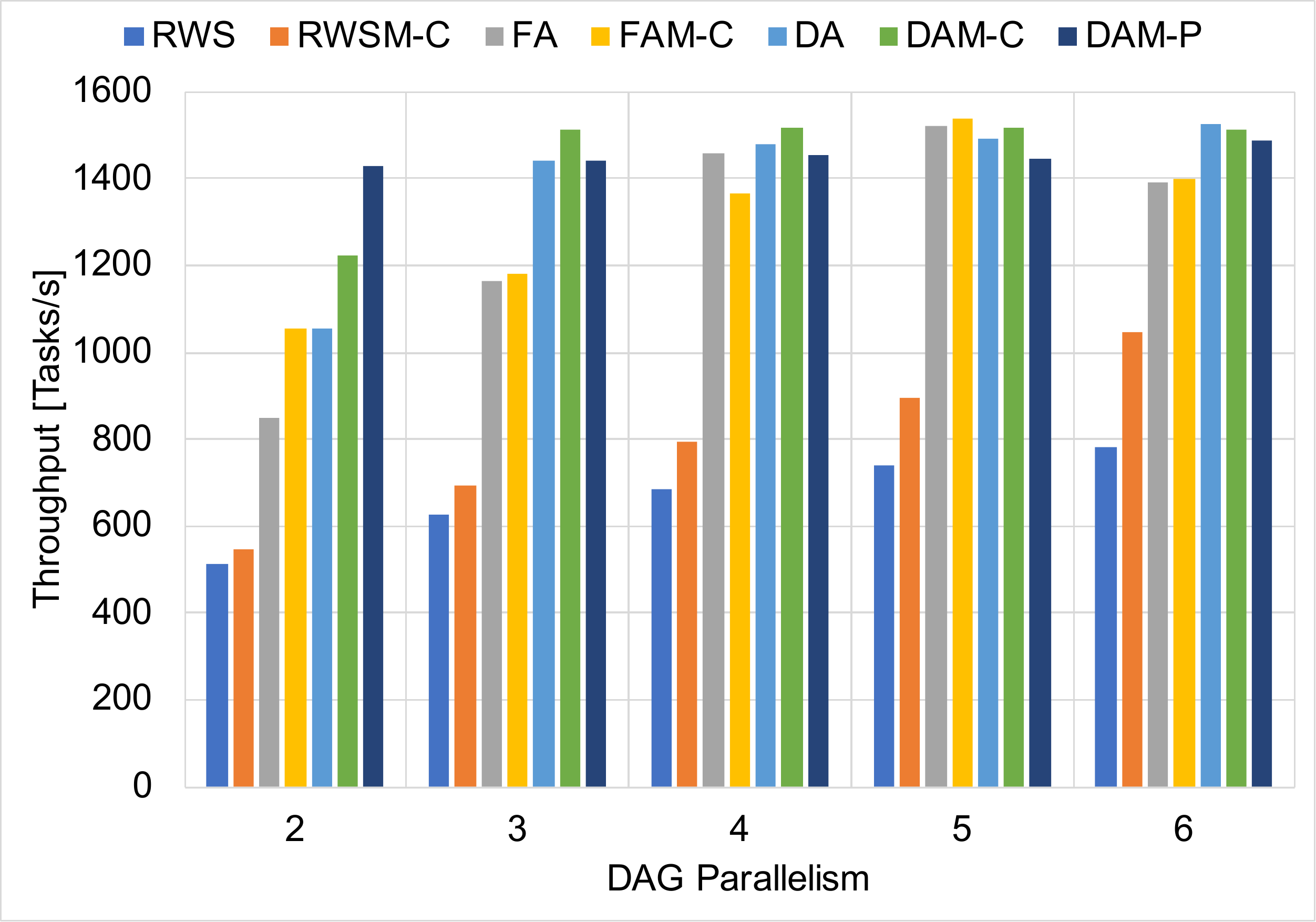}}
\subfigure[Stencil]{\label{fig_dvfs_sten} \includegraphics[width=0.32\textwidth]{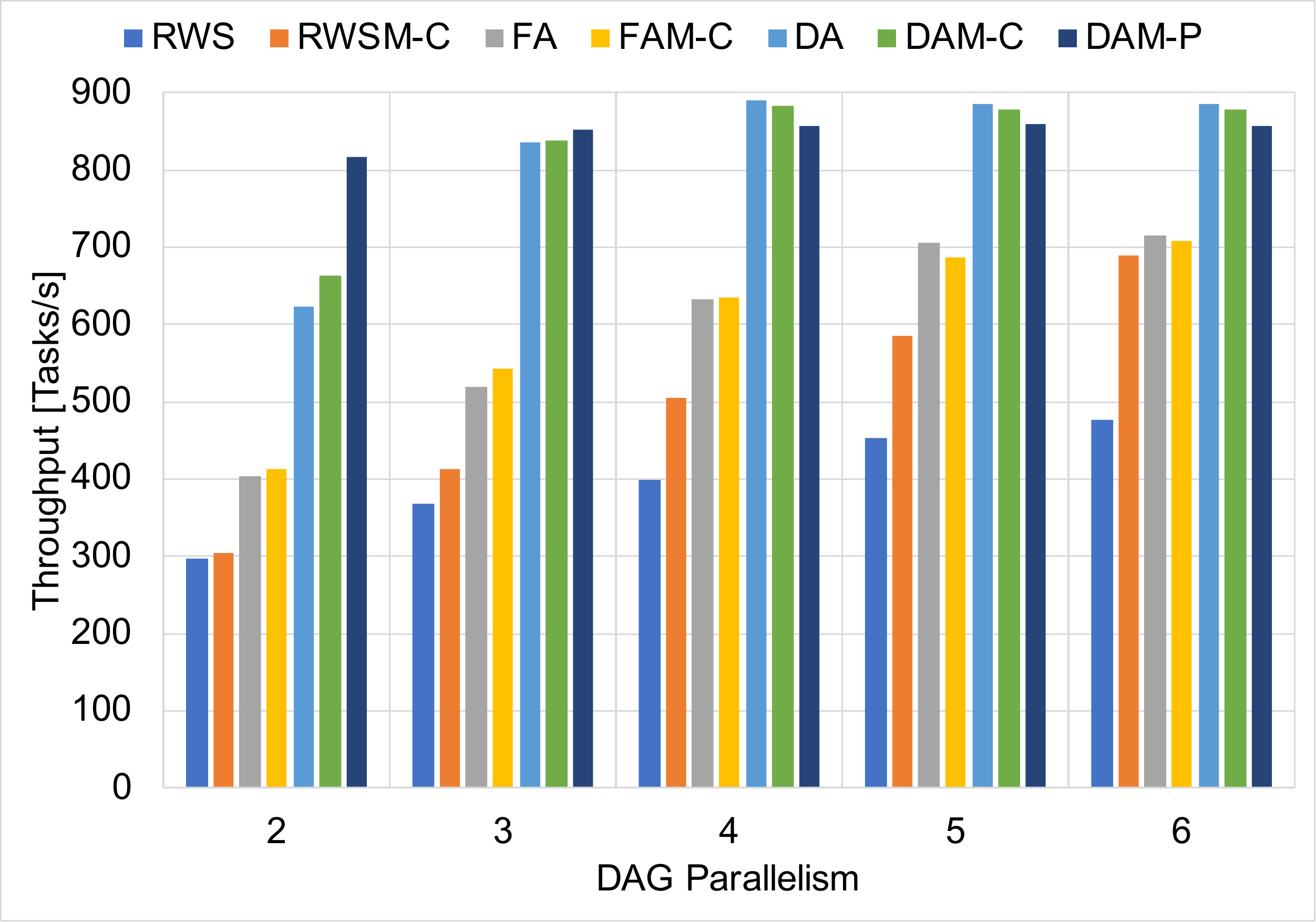}}
\caption{The performance impact of DVFS and comparison between different schedulers.} 
\label{fig:dvfs}
\end{figure*}

To analyze the response to DVFS, we induce periodic frequency changes in the TX2 Denver cluster alternating between the highest and the lowest frequency (2035 MHz and 345 MHz, respectively) with a 10s period for a full cycle (i.e.~5s+5s).
The performance impact on the different schedulers due to such event is shown in Figure~\ref{fig:dvfs}.
We note that DA, DAM-C and DAM-P are all more resilient to DVFS than other schedulers.
For instance, with the \textit{copy} benchmark (shown in Figure~\ref{fig_dvfs_copy}), DAM-C achieves roughly 2.2$\times$ and 1.9$\times$ average performance speedup relative to RWS and RWSM-C across different degrees of DAG parallelism.
Additionally, DAM-C provides an average 17\%  and 12\% throughput improvement over FA and FAM-C, respectively.
Nevertheless, DAM-P performs better than DA and DAM-C when parallelism is low since it tries to minimize the execution time of priority tasks by selecting the fastest execution place irrespective of parallel cost. 
Consequently, it is more likely to exploit multiple cores per task.
In contrast, the DAM-C variant conservatively chooses lower widths during interference events. 
This is due to reducing the parallel cost by excluding the cores with perturbed performance. 
At low parallelism, reducing parallel cost leads to a sub-optimal schedule due to increased idleness. In such cases, DAM-P performs better as it targets highest parallel performance for priority tasks to compensate for the low parallelism. 
\subsection{Sensitivity Analysis}
\label{Perf_SA}
\begin{figure}[h]
\centering
\includegraphics[width=0.34\textwidth]{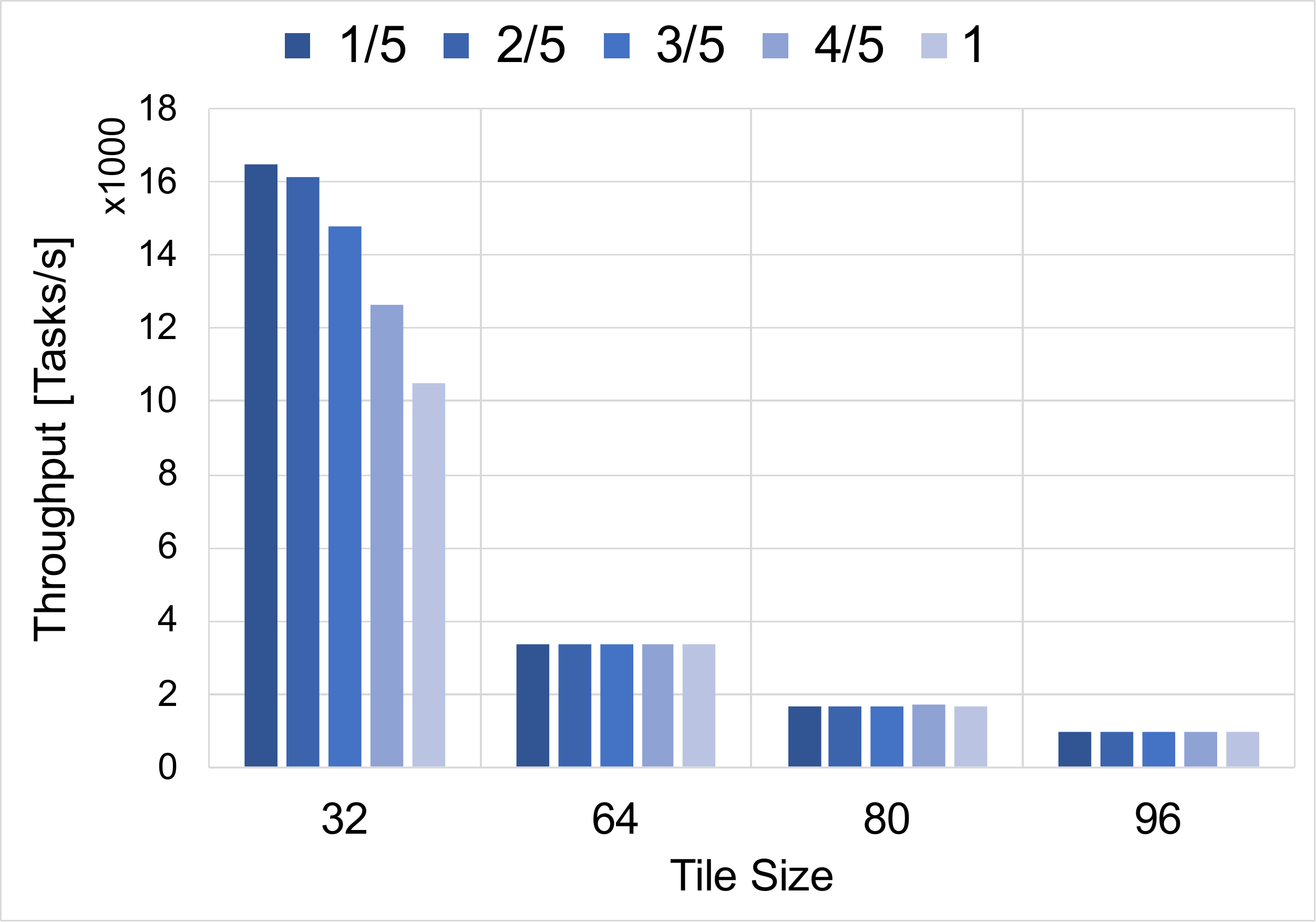}
\caption{Sensitivity analysis of data tile size and weight ratio of updating PTT.}  \vspace{-3mm}
\label{fig:Sensitivity}
\end{figure}

Figure~\ref{fig:Sensitivity} presents the sensitivity analysis of the weighted update strategy of PTT discussed in section~\ref{ptt_impl} and the performance impact from different tile size, for \textit{matrix multiplication}. 
The legends in Figure~\ref{fig:Sensitivity} demonstrate the weight ratio of new execution time.
For instance, 2/5 suggests that the new execution time is assigned a weight of 2 and the old execution time in PTT is assigned a weight of $5-2=3$. 
Since the L1 data cache size is different for A57 (32KB) and Denver (64KB) cores, we test different tile sizes to understand the impact of tile size on throughput. 
Note that with the tile size of 32 the working set of a task fits in both the A57 and the Denver L1 data caches while with the tile size of 64 and 80 it only fits in the Denver L1 data cache. The tile size of 96 indicates the case when the working set will only fit in the L2 data cache (2MB) of each cluster.
It can be seen that the performance is only impacted by the weight ratio of updating strategy when the tile size is 32, which shows that 1/5 in this case is the best.
The throughput breakdown between the best and worst is around 36\%.
When the tile size increases, the weight ratio has much less impact on performance and the throughput keeps stable.
Therefore, we select 1/5, i.e. 1:4, for the updating PTT in this paper.

\subsection{Interference on Applications}
\label{Perf_Haswell}


\begin{figure*}[t]
\centering
\subfigure[K-means Clustering]{\label{fig_kmeans_interference} \includegraphics[width=0.32\textwidth]{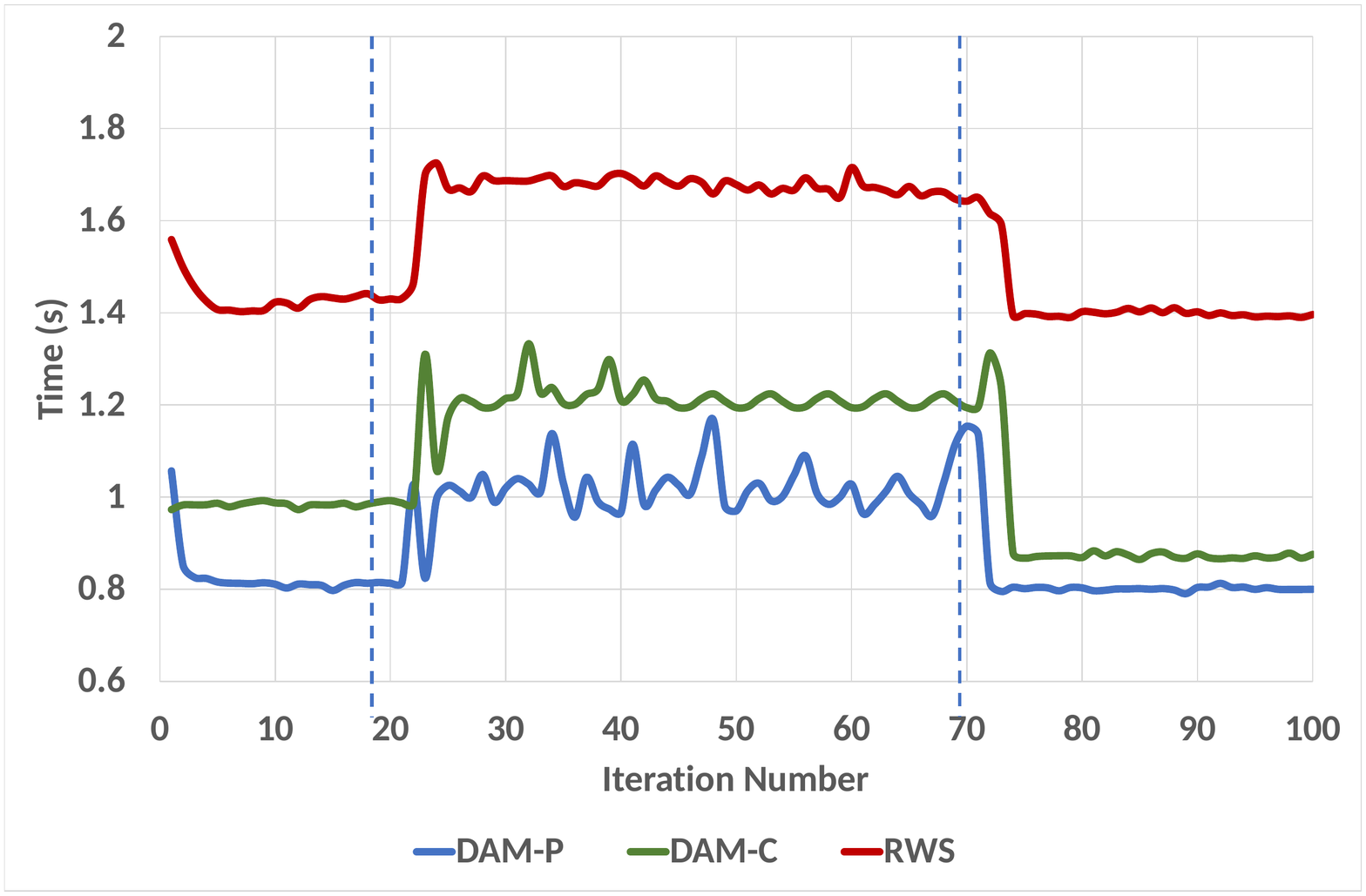}}
\subfigure[RWS]{\label{fig_rws_response} \includegraphics[width=0.32\textwidth]{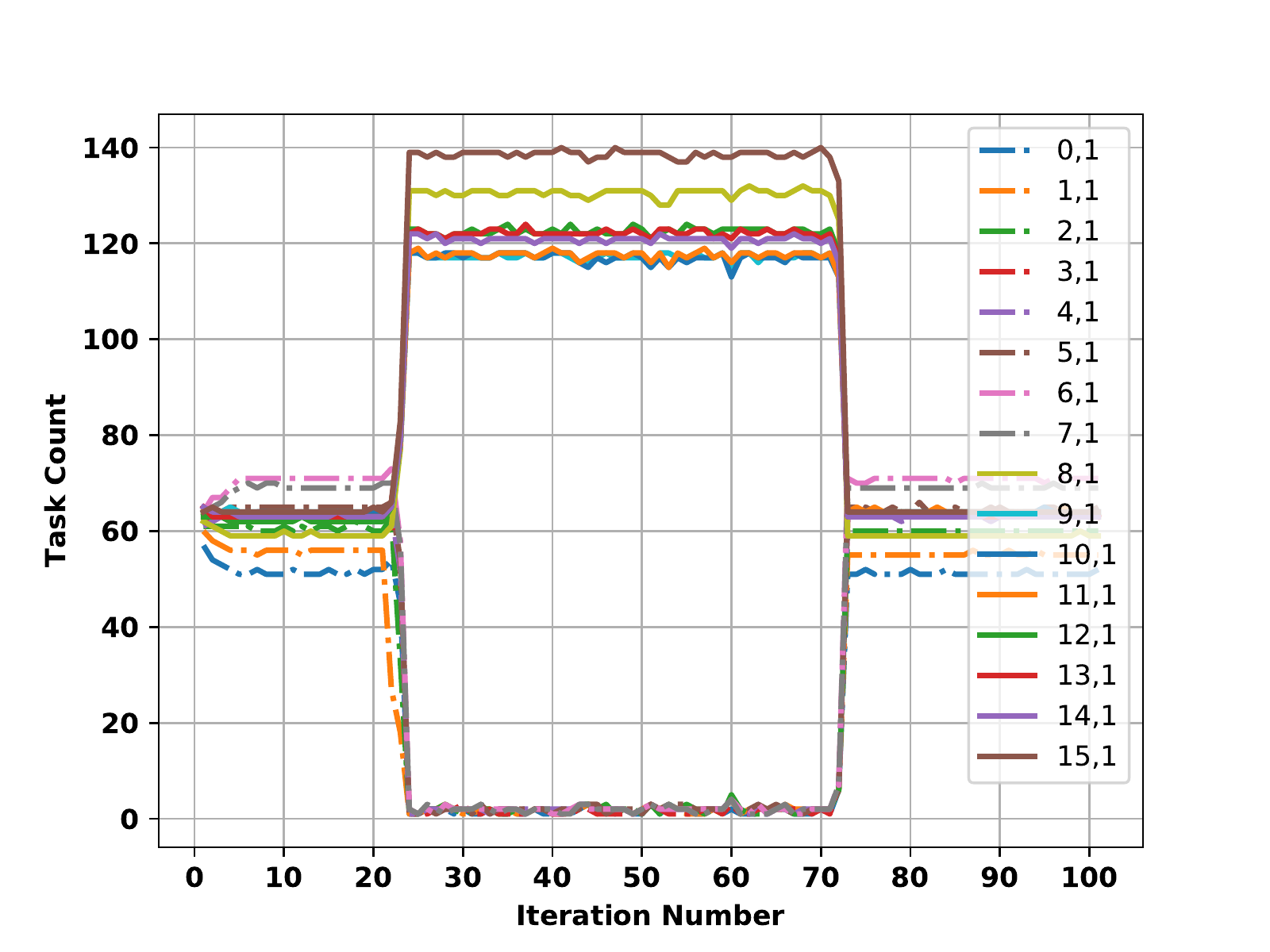}}
\subfigure[DAM-P]{\label{fig_ptt_response} \includegraphics[width=0.32\textwidth]{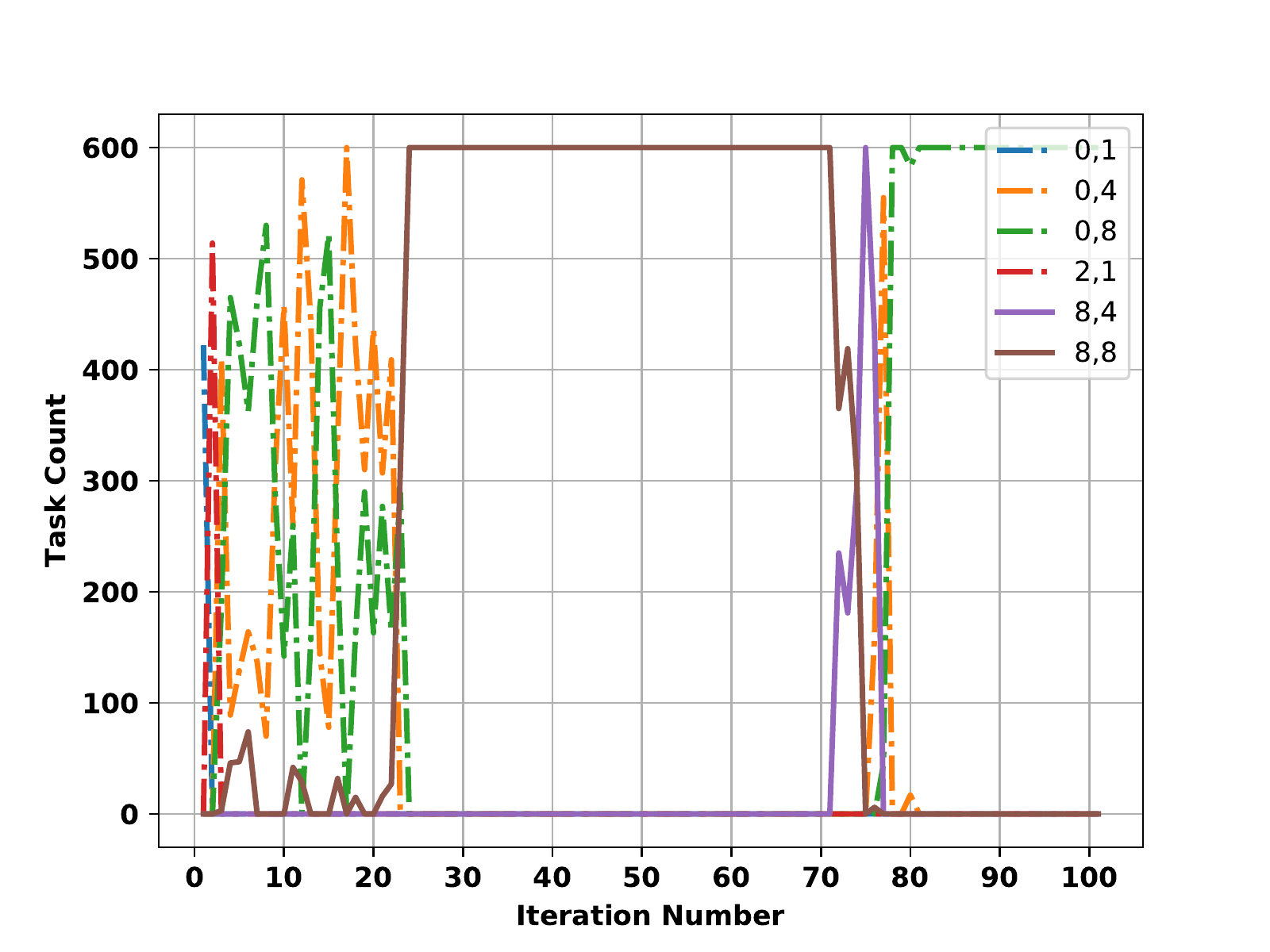}}
\caption{Performance of K-means clustering on 16-core Haswell (a), and high priority resource selection during interference on socket 0 across K-means clustering iterations (20-70) for DAM-P and RWS ((b) \& (c)). The dotted lines represent the socket 0 partitions}  
\label{fig:interference_respose}
\end{figure*}

We now discuss the performance of \textit{K-means} and \textit{2D Heat} on a 10-core dual-socket Haswell platform. In these experiments, we run the interfering app on a single of the two sockets.
One of the challenges that needs to be addressed is that a simple model like the PTT may not have enough training data within a single iteration to detect interference. In other words, for the 20 cores of this configuration, there are many resource partition choices to exhaust. Hence, the co-running application starts a few iterations after the start ensuring a reasonable window for training, since both codes are iterative. 
In K-Means, we map the loop partitions to dynamically scheduled tasks and assign the high priority to the task containing the largest work unit. In the case of 2D Heat, the critical tasks are those that perform boundary point communication.  
Figure
~\ref{fig_kmeans_interference} plot the execution time ($Y$-axis) across different iterations ($X$-axis) for K-Means.
We drop the FA and FAM-C schedulers because the Intel Haswell platform is statically symmetric. 
The background interference interval is marked by the blue dotted lines. On average, DAM-P exhibits the best performance during interference. However, due to the limited accuracy of system clocks and changing conditions during time measurement, variability arises across iterations.  
The analysis in Figure~\ref{fig:interference_respose} (b) and (c) provides insight into the differences in the context of K-means. Each curve here represents the number of tasks scheduled using the corresponding execution place in the K-means application. Figure~\ref{fig_rws_response} shows that RWS selects tasks for scheduling on the interference socket (marked by dotted lines), but also it shows load-imbalance mostly due to intra-task interference and resource over-subscription (140 tasks on core 15, 130 on core 8 and around 120 on the rest). Figure~\ref{fig_ptt_response} shows that DAM-P prefers to mold the tasks on the 8 cores of socket 1 during the event, thus minimizing inter and intra process interference for high priority \textit{K-means} tasks.

\begin{figure}[h]
\centering
\includegraphics[width=0.3\textwidth]{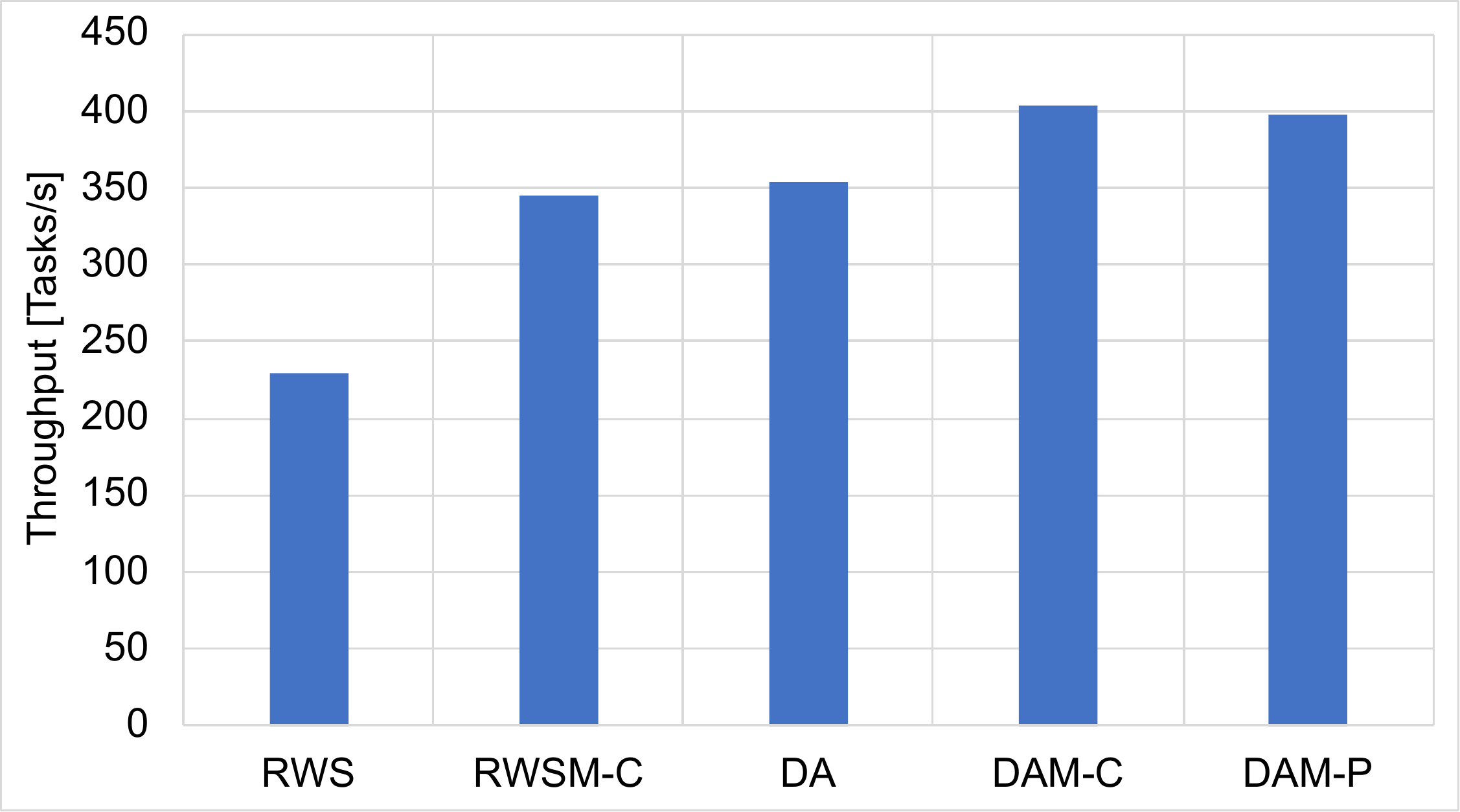}
\caption{Performance comparison of distributed 2D Heat using different schedulers.} 
\vspace{-3mm}
\label{fig:MPI_heat}
\end{figure}

Finally, we show the performance a distributed memory DAG-based 2D \textit{Heat} stencil application on 4 ``haswell'' nodes (totaling 80 cores) as shown in Figure~\ref{fig:MPI_heat}. The interference \textit{matrix multiplication} kernel is executed on 5 cores of a single socket of node 0. It can be seen that the performance of Dynamic Asymmetric Schedulers in general are better than other schedulers. Even though communication tasks utilize a single core at a time, which is an inherent nature of message passing, moldability reduces resource contention and provides a benefit since sharing CPU caches can have a significant impact on MPI communication \cite{proquedis-mpi-cache-cluster12}. This explains the higher throughput of DAM-C and DAM-P compared to DA. Additionally, DAM-C achieves 76\% and 17\% throughput increases compared to RWS and RWSM-C, respectively. 

\section{Related Work}\label{RelatedWork}
Performance variability due to interference is highly evident as it may arise from on-chip variations \cite{process1,process2}, network activity \cite{patki-sc19,network2}, I/O traffic \cite{io1,io2}, etc. 
To study performance variability, Ates et al.~\cite{ates-icpp19} introduce a performance anomaly generator for the major HPC subsystems that assesses the performance resilience of applications to different variability sources.
Works that try to mitigate interference can broadly be classified into two groups: those that deal with interference at the cluster level and those that deal with interference at the node level.

System noise at the cluster level has been thoroughly analyzed on large-scale architectures. For example, Hoefler et al.~use LogGPS simulations to get insights into the scaling of applications in noisy environments~\cite{system-noise-hoefler}. In ~\cite{skinner-wcs05}, inter and intra-application resource contention are classified as sources of performance variability. On the runtime systems' side, a real-time monitoring component that assists the runtime scheduler in its decision-making process and adaptation is developed as part of the AllScale toolchain~\cite{allscale}. By collecting information across a cluster, the AllScale runtime can tune parameters such as thread counts or DVFS. However, the collected information is too coarse to address either asymmetry or interference by influencing task scheduling.

At the node level, the idea of using system behavior observations has been proposed to influence the design of OS scheduling~\cite{os-obs-knau} to reduce the interference caused by shared resources in chip multiprocessors. 
Zhuravlev~\cite{zhuravlev-asplos10} propose a new scheduling algorithm that mitigates the effects of shared resource contention. This scheduler, called Distributed Intensity Online (DIO), collects miss rates online for all applications and schedules applications to minimize performance impact.
%
At the runtime level, Chronaki et al.~\cite{chronaki-tpds17} introduce various schedulers, including CATS, based on the idea of steering critical tasks to statically faster cores. To evaluate CATS the authors introduce the dynamic Heterogeneous Earliest Finish Time (dHEFT) algorithm as a reference to evaluate CATS. HEFT (Heterogeneous Earliest Finish Time) is a static scheduling algorithm that assigns each task to the processor that will finish its execution at the earliest possible time~\cite{topcuoglu-tpds02}. 
dHEFT uses the same principles as HEFT but instead of knowing the load of tasks prior to scheduling, discovers them at runtime. 
While these schedulers can improve the execution time of task-DAGs in which tasks have diverse behaviors, they have a few limitations. First, none of them are able to avoid resource over-subscription and adapt to dynamic asymmetry. And second, all of them are based on the notion of only two static performance classes, i.e.~big and LITTLE. Not only are our Dynamic Asymmetry schedulers able to model the performance of all cores without prior assumptions on the hardware, but they can also exploit this information to mitigate the impact of interference.   

\section{Conclusion} 
In this paper we have explored techniques for effective scheduling of parallel applications in the presence of dynamic performance asymmetry. 
Our findings indicate that random work stealing schedulers are not effective because they do not leverage information about task priority nor on the capability or state of the underlying resources. Schedulers that have a fixed notion of platform asymmetry perform better than random work stealing schedulers but still leave a lot of room for improvement because they do not posses knowledge about dynamic changes in the execution environment. Our proposed dynamic asymmetry aware schedulers
schedule high-priority tasks around interference and enable moldable execution of tasks resulting in a improvement over traditional scheduling approaches, not only during user-level background activity but also during DVFS. The proposed schedulers can detect and adapt to interference by using an online performance model based on a performance tracing table. 
All in all, we believe that combining user-level approaches such as \scheduler{} (DAS) with system-level approaches is a significant step to achieve tolerance to the increasingly challenging problem of unpredictable system interference.  

\section{Acknowledgment} \label{conclusion}
The computations/data handling were enabled by resources provided by the Swedish National Infrastructure for Computing (SNIC) at C3SE partially funded by the Swedish Research Council through grant agreement no. 2016-07213. The research leading to these results has received funding from
the European Union’s Horizon 2020 Programme under the LEGaTO Project (www.legato-project.eu), grant agreement no 780681.

\bibliographystyle{ACM-Reference-Format}
\bibliography{Jing3,jrlib3}

\end{document}